 \documentclass[preprint,review, 11pt]{elsarticle}

\usepackage{framed,multirow}
\usepackage{rotating}
\usepackage{enumitem}
\usepackage{lscape}
\usepackage{booktabs}
\usepackage{amssymb}
\usepackage{amsmath}
\usepackage{amsthm}
\usepackage{latexsym}
\usepackage{float}
\usepackage{algorithm}
 \usepackage{algpseudocode}
\usepackage{setspace}
\usepackage{url}
\usepackage{xcolor}
\usepackage{caption}
\usepackage{subcaption}
\usepackage[margin=1.5cm]{geometry}

\usepackage{tabularx} 
\usepackage{booktabs} 
\usepackage{graphicx} 
\usepackage{hyperref}

\makeatletter
\def\ps@pprintTitle{%
     \let\@oddhead\@empty
     \let\@evenhead\@empty
     \def\@oddfoot
       {\hbox to \textwidth%
        {\ifnopreprintline\relax\else
        \@myfooterfont%
         \ifx\@elsarticlemyfooteralign\@elsarticlemyfooteraligncenter%
           \hfil\@elsarticlemyfooter\hfil%
         \else%
         \ifx\@elsarticlemyfooteralign\@elsarticlemyfooteralignleft%
           \@elsarticlemyfooter\hfill{}%
         \else%
         \ifx\@elsarticlemyfooteralign\@elsarticlemyfooteralignright%
           {}\hfill\@elsarticlemyfooter%
         \else%
               Preprint submitted to \ifx\@journal\@empty%
                 arxiv.org%
            \else\@journal\fi\hfill\@date\fi%
         \fi%
         \fi%
         \fi%
         }%
       }%
     \let\@evenfoot\@oddfoot}
\makeatother

\date{June 2026}

\begin{document}

\begin{frontmatter}

\title{An unsupervised kernel norm monitoring for fault detection in a time series photovoltaic system.}

\author[1]{Victoria Jorry}\corref{cor1}
\cortext[cor1]{Corresponding author}
\ead{Victoria.Jorry@lut.fi}
\author[1]{Zina-Sabrina Duma}
\author[1]{Satu-Pia Reinikainen}
\author[1]{Heikki Haario}
\author[1]{Lassi Roininen}
\address[1]{LUT University, Yliopistonkatu 34, Lappeenranta 53850, Finland}

\begin{abstract}
Grid-connected photovoltaic systems (GCPVS) are generally robust but remain susceptible to faults that can compromise energy conversion efficiency or raise safety concerns. Promptly and automatically detecting such anomalies is therefore essential for maintaining system reliability and performance. However, in practice, labeled fault data are rarely available in real-world deployments, which limits the applicability of supervised approaches.
Conventional unsupervised baseline models, including a one-class support vector machine (OCSVM), isolation forest (iForest), and local outlier factor (LOF), are trained on normal operation data and assign anomaly scores reflecting how closely new observations resemble that baseline. Although these methods already accommodate non-linear behavior to varying degrees, kernel-based formulations offer further flexibility in shaping the decision boundary; however, tuning the kernel hyperparameters ordinarily requires some prior knowledge of the fault regime. We overcome this limitation by proposing kernel-based norm monitoring (KNM), a non-linear, unsupervised, window-based fault-detection method designed for continuous processes. Although the paper focuses on the GCPVS as a case study, KNM is a general-purpose monitoring framework applicable to a wide range of industrial processes.
Using the Grid-connected PV System Faults (GPVS-Faults) dataset operating in intermediate power point tracking (IPPT) mode, KNM is evaluated in two fault scenarios, sensor faults and partial shading, against three benchmark techniques: OCSVM, iForest, and LOF. KNM achieves up to 99.1\% and 98.3\% accuracy on the two fault scenarios, respectively, using the Cauchy kernel, compared to 93.5\% for the best-performing benchmark. The method is interpretable, and variable contribution plots are proposed to support fault identification.\end{abstract}

\end{frontmatter}

\section{Introduction}

As a renewable and sufficient energy source, photovoltaic (PV) systems have experienced tremendous growth and have contributed significantly to transforming the world into a clean, green energy environment \cite{ali2025advancements}. PV assemblies are vulnerable to various irregularities and failures, which can reduce system efficiency and power generation and even pose safety risks \cite{mellit2018fault}. Unidentified faults may be the cause of lower-energy outputs, higher maintenance costs, and additional safety risks \cite{amiri2024faults}. Hence, detecting them in solar energy systems is essential to guarantee reliability, effectiveness, and operational longevity \cite{khalil2024novel}. Anomaly detection includes routine monitoring and assessment of PV system components to detect and diagnose deviations from typical operation. 

Depending on which part of the PV system is disturbed, the anomalies can be classified into two groups: the Direct Current (DC) side, positioned before the inverter, and the Alternating Current (AC) side, which is the inverter's output. The PV system suffers from energy output due to DC-side anomalies; common failure modes include electrical mismatches between modules, localized hotspot formation, insulation degradation, and open or short-circuit conditions in the DC/DC converter \cite{harrou2024automatic}. Likewise, the AC side tends to experience issues like total blackouts and unexpected grid irregularities. Operators can promptly locate and resolve problems on both DC and AC sides of PV systems by using modern anomaly-detection algorithms to detect departures from typical operation behavior \cite{vai2022study}.

In recent decades, extensive research has focused on developing several fault detection methods in photovoltaic system monitoring \cite{harrou2019unsupervised,sepulveda2025artificial,araneo2009emc}. These methods can be classified into two groups: model-based and data-based approaches, commonly applied for fault detection in PV systems \cite{toche2024comprehensive}. Model-based approaches rely on mathematical or physical models to describe the expected behavior of PV systems \cite{sepulveda2025artificial}, developed based on an understanding of the underlying physics and engineering principles governing PV systems \cite{skomedal2021robust}. Comparing actual system measurements with the expected model predictions can help identify deviations or anomalies, indicating abnormal behavior \cite{garoudja2017statistical}. Model-based approaches offer valuable insights into the root causes of anomalies and can aid in system optimization and maintenance, but they require accurate and detailed knowledge of system dynamics and may struggle with complex and non-linear behaviors \cite{harrou2024automatic}. Several methods have been developed for detecting anomalies in PV plants, such as diode-based models \cite{huang2022parameter}, and the application of a Kalman filter \cite{al2023multiple}. The success of model-based techniques hinges on obtaining accurate analytical models, which can be a time-consuming and challenging task for large-scale PV systems \cite{harrou2017model}.

On the other hand, data-driven approaches rely on historical PV system data to detect abnormalities \cite{ning2021data}. These frameworks use machine learning and statistical methods to analyze data and identify patterns that distinguish normal from abnormal behavior \cite{xiong2025research}. Data-based approaches do not rely on explicit models of the system, but rather learn from the available data to establish normal operating patterns \cite{sepulveda2025artificial}.  Several multivariate statistical monitoring methods based on dimensionality reduction techniques, such as principal component analysis \cite{amaral2021fault,mandal2020pv}, improved independent component analysis \cite{harrou2025anomaly}, and partial least squares \cite{taghezouit2024model}, have been considered within data-driven methods for monitoring PV systems. For example, a recent study in \cite{bouyeddou2022improved} introduced an efficient approach to identify anomalies in grid-connected photovoltaic plants using latent variable regression methods along with the triple exponentially weighted moving average anomaly detector.  In another study \cite{chen2017adaptive}, an anomaly detection scheme for PV systems was proposed that uses multiple meters to measure different output signals and leverages their correlations via a vector autoregressive model.

In recent years, there has been a growing interest in exploring various machine learning approaches to enhance anomaly detection in PV systems \cite{amiri2024faults}. They provide flexibility for managing diverse data sources within PV systems and can uncover intricate relationships, enabling the identification of abnormalities or faults that are not easily detectable with traditional techniques \cite{fazai2019machine}. Related studies include artificial neural networks \cite{khalil2024novel}, decision trees \cite{benkercha2018fault}, combinations of support vector machines with naive Bayes and k-nearest neighbors classifiers \cite{eskandari2020line}, random forests \cite{chen2018random}, probabilistic neural networks \cite{garoudja2017statistical}, and gradient boosting variants such as CatBoost, LightGBM, and XGBoost \cite{adhya2022performance}. While these supervised methods have shown high accuracies, they typically require labeled data for training, which may not be available for all fault categories or for previously unseen fault types \cite{harrou2025anomaly}. 

This limitation has motivated the growth of semi-supervised and unsupervised alternatives, which rely primarily on normal operational data for training while maintaining competitive detection performance \cite{wang2023self}. In real PV systems, labeled data are difficult to obtain due to their scarcity, unpredictability, and fault diversity \cite{scholkopf2001estimating}. As a result, recent research has increasingly focused on semi-supervised and unsupervised methods that mainly train on normal data and have shown promise in capturing subtle anomalies and improving predictive maintenance and system reliability \cite{liu2025vision}. The systematic benchmarking study in \cite{harrou2026semi} evaluated one-class support vector machine (OCSVM), isolation forest (iForest), local outlier factor (LOF), and elliptic envelope methods for fault detection in GCPVS across multiple operating modes. The present work builds directly on that benchmarking framework: we compare against OCSVM, iForest, and LOF under IPPT conditions.

Harrou et al. \cite{harrou2024automatic} proposed a semi-supervised procedure that combines variational autoencoders (VAE) with various anomaly detection algorithms, including iForest and OCSVM, for fault detection in GCPVS. The proposed approach achieved up to 92.99\% accuracy in the Intermediate Power Point Tracking (IPPT) operating mode for identifying different fault scenarios. Furthermore, in \cite{harrou2026semi}, a systematic mode-aware benchmarking of semi-supervised anomaly detection methods (OCSVM, iForest, LOF, and Elliptic Envelope (EE)) for fault detection in GCPVS was presented. On the other hand, deep learning methods have gained popularity as they can automatically identify complex and hidden patterns within data \cite{dairi2020short,mansouri2021deep,duranay2023fault,parvin2025photovoltaic}. Some of the related studies in this category include a convolutional neural network (CNN) model paired with a meta-heuristic algorithm \cite{seghiour2023deep}, a one-dimensional CNN \cite{sridharan2025convolutional}, a Deep CNN \cite{wang2019novel}, a KNN-based random subspace (RS) ensemble classifier \cite{swarna2022knn}, and a stacked sparse autoencoder (SAE) \cite{qi2017stacked,liu2021fault}.

Among unsupervised statistical monitoring approaches, kernel methods occupy a particularly promising position. By implicitly mapping process observations into a high-dimensional reproducing kernel Hilbert space (RKHS) through a positive definite kernel function, these methods can capture non-linear dependencies among process variables that linear methods such as principal component analysis (PCA) miss entirely \cite{pilario2019review}. In fault detection, this translates to a monitoring statistic that is sensitive to structural changes in the process distribution --- including subtle distributed faults that do not produce clear outliers in the original measurement space \cite{huang2014high}. Despite this potential, the application of kernel methods to unsupervised fault detection in photovoltaic systems remains limited, partly because tuning the kernel parameters without labeled fault data is a non-trivial problem. The present work addresses this gap directly.

Despite substantial progress in PV fault detection, many reported methods still depend on supervised models that require high-dimensional sensor fusion and are trained with labeled fault data, which is rarely available in practical operating environments. In addition, several recent studies demonstrate strong detection performance only under MPPT conditions \cite{khan2023data,teta2024fault}, where persistent control perturbations increase signal variability. IPPT, on the contrary, is characterized by regulated intermediate-power operation, reduced short-term dynamics, more compact feature distributions, and modified control loop behavior \cite{bakdi2021real}. Owing to these differences, models, decision boundaries, and anomaly statistics developed for MPPT operation cannot be presumed to generalize directly to IPPT. Fault detection under IPPT should therefore be treated as a separate modeling problem requiring dedicated evaluation. These gaps motivate the development of efficient, unsupervised learning-based approaches that leverage standard fault-free training data, accommodate the IPPT operating mode, and provide interpretable outputs for engineering decision support. In response, this study proposes an unsupervised fault detection framework named kernel-based norm monitoring (KNM), which is evaluated under both univariate and multivariate sensor configurations, providing systematic evidence on the minimum sensor requirements for reliable fault detection under IPPT operating conditions. The main contributions of this study are summarized as follows:

\begin{enumerate}[label=(\roman*)]
    \item An adaptive kernel-based unsupervised fault detection framework is proposed for photovoltaic systems operating under the IPPT regime. The framework dynamically optimizes kernel parameters using a variance-maximization criterion, eliminating the need for labeled fault data during model development
    \item A KNM statistic based on the Frobenius norm of the centered kernel feature matrices is introduced to quantify structural changes in nonlinear process behavior within a Reproducing Kernel Hilbert Space (RKHS). The proposed statistic enables fault detection through deviations from fault-free operating conditions.
    \item The effectiveness of low-dimensional sensor monitoring is investigated through both univariate and multivariate configurations, demonstrating that reliable fault detection can be achieved using a limited subset of photovoltaic measurements rather than full-system instrumentation
    \item An interpretability layer based on per-variable kernel norm decomposition, enabling identification of the process variables responsible for a detected fault. This provides physically meaningful diagnostic outputs that are consistent with operator decision-support requirements --- addressing a known limitation of black-box unsupervised anomaly detectors.
    \item The proposed framework is validated using the publicly available GPVS-Faults dataset under IPPT operation and compared against established unsupervised anomaly detection methods, including OCSVM, iForest, and LOF.
\end{enumerate}

The remainder of this article is organized as follows: Section~\ref{sec:1} presents the simulated and real process data and the base mathematical methods used; Section~\ref{sec:results} presents the results of the proposed KNM framework, and Section~\ref{sec:conclusion} concludes the findings.

\section{Materials and methods}
\label{sec:1}
This section describes the dataset and methods used to estimate kernel parameters for fault detection in time-series data, such as those from a solar photovoltaic system. The methods first divide the real-time-measured signals into overlapping windows using a sliding-window strategy. Frequency domain features are subsequently extracted from each window using the Fast Fourier Transform (FFT), and the resulting feature vectors are used to construct kernel representations for unsupervised fault detection and monitoring.

\subsection{Dataset description}
\label{sec:2}

Two case studies are considered to evaluate the proposed framework. The first consists of a synthetic nonlinear process designed to provide a controlled environment for illustrating the behavior of the proposed monitoring statistic under gradual fault development. The second employs the publicly available GPVS-Faults dataset, which contains experimentally generated photovoltaic faults under IPPT operation. The combination of simulated and experimental case studies enables both methodological validation and practical assessment of the proposed approach.
    
\subsubsection{Simulated dataset description}
\label{sec:2}
To evaluate the proposed KNM framework under controlled and reproducible nonlinear operating conditions, a synthetic dataset is generated consisting of $n_\text{normal}$ healthy samples followed by $n_\text{faulty}$ faulty samples. Let 
\[
\mathbf{z}_t \in \mathbb{R}^d, \quad t = 1, \ldots, n_{\text{normal}}+n_{\text{faulty}},
\]
denote a latent random vector sampled from a multivariate normal distribution 
\[
\mathbf{z}_t \sim \mathcal{N}(\mathbf{0}, \boldsymbol{\Sigma}_\text{latent}), \quad \text{where } \boldsymbol{\Sigma}_\text{latent}(i,j) = 
\begin{cases}
1.0 & \text{if } i = j, \\
0.6 & \text{if } i \neq j.
\end{cases}
\]

This covariance structure introduces moderate cross-variable correlation, emulating the interdependencies typically observed between electrical variables in a photovoltaic system.
    
To create nonlinear process dynamics, each latent variable is transformed according to its index:
    \[
    x_{t,j} = 
        \begin{cases}
            z_{t,j}^2 + 0.5\, z_{t,j}, & j \equiv 1 \pmod{3} \quad \text{(quadratic)}, \\
            \sin(z_{t,j}),          & j \equiv 2 \pmod{3} \quad \text{(sinusoidal)}, \\
            \exp(-z_{t,j}^2),       & j \equiv 0 \pmod{3} \quad \text{(Gaussian bump)}.
        \end{cases}
    \]
    
The resulting dataset, therefore, contains a mixture of quadratic, sinusoidal, and Gaussian-shaped nonlinear relationships while preserving the underlying correlation structure. After the healthy operating period, a fault is introduced at $t = n_\text{normal} + 1$. The fault affects only the first two variables, while all remaining variables remain unchanged. Let $\delta_f > 0 $ denote the fault magnitude parameter. The faulty observations are generated as
    \[
    x_{t,1}^{(f)} = x_{t,1} + \delta_f x_{t,1}^2,
    \]  and
    \[
    x_{t,2}^{(f)} = x_{t,2} + \delta_f \sin(\pi x_{t,2}),
    \] for
    \[
    t = n_{\text{normal}} + 1, \dots, n_{\text{normal}} + n_{\text{faulty}}.
    \]
    
These modifications introduce nonlinear distributional changes and progressive departures from the healthy operating regime while leaving the majority of process variables unaffected. The transformed observations are assembled into the vector 
    $\mathbf{x}_t = (x_{t,1}, \ldots, x_{t,d})^\top \in \mathbb{R}^d$, 
    and the full dataset is collected into the matrix 
    $\mathbf{X} \in \mathbb{R}^{N \times d}$, where $N = n_\text{normal} + n_\text{faulty}$.
    
By varying the fault magnitude parameter $\delta_f$, the severity of the fault can be systematically controlled, enabling quantitative evaluation of fault sensitivity, false alarm behavior, and detection delay. The resulting dataset, therefore, provides a challenging benchmark for assessing the ability of kernel-based monitoring methods to detect localized nonlinear faults that may not be readily identified using conventional linear MSPC approaches.
    

\subsubsection{Process data}
The dataset used in the research includes real-time fault scenarios from GCPVS operating in MPPT and IPPT modes \cite{bakdi2020gpvs}, as shown in Fig.~\ref{fig:Blockdiag}. The data comes from a series of experiments involving seven different types of faults, labeled F1 to F7 (summarized in Table~\ref{tab:1}), including inverter issues, feedback sensor failures, and grid anomalies. The faults vary in type and location to provide a comprehensive analysis. Each fault is manually introduced in multiple independent experiments, with each experiment lasting approximately 10 to 15 seconds. The faults are typically introduced between the 7\textsuperscript{th} and 8\textsuperscript{th} seconds of the experiment. The sampling time for the acquisition of fault-free and defective data is \( T_s \approx 9.999 \, \mu s \). The experiments were conducted with varying environmental conditions, including changes in temperature and solar irradiance, which can introduce noise into the measurements. %

Each data file contains time-series measurements of various electrical variables, including PV array current (Ipv), PV array voltage (Vpv), and DC voltage (Vdc), as well as three-phase current and voltage measurements (Ia, Ib, Ic, Va, Vb, Vc). Additional features, such as positive-sequence estimated current magnitude and frequency, as well as voltage magnitude and frequency, are also included in the dataset. The data files are labeled by fault type and operation mode (MPPT or IPPT), with fault-free scenarios included for training the fault-detection models. The dataset's high frequency and noise pose challenges for fault detection, as does the presence of MPPT/IPPT controllers that can mask low-magnitude faults.

However, since the majority of recent work has focused on MPPT operation with all 13 variables \cite{solorzano2013automatic,khan2023data,li2021fast}, where constant disruptions introduce greater signal variability. Conversely, IPPT operates under regulated intermediate-power conditions with reduced short-term dynamics, narrower feature distributions, and altered control-loop behavior \cite{harrou2026semi}. As a result, detection boundaries and anomaly score statistics learned under MPPT cannot be assumed to transfer reliably to IPPT, making IPPT analysis non-trivial and prompting mode-aware modeling. Thus, in this study, we evaluate our proposed framework using only four sensitive variables (Ipv, Vf, Vdc, and Vpv) to identify F2 and F4 faults under IPPT mode. 

\begin{table}[H]
\centering
\caption{Realistic injected faults in the GPV system.}
\label{tab:1}
\begin{tabular}{llp{11cm}}
\toprule
\textbf{Fault} & \textbf{Type} & \textbf{Description} \\
\midrule
F1 & Inverter fault & Complete failure in one of the six IGBTs \\
F2 & Feedback Sensor fault & One phase sensor fault 20\% \\
F3 & Grid anomaly & Intermittent voltage sags \\
F4 & PV array mismatch & 10 to 20\% nonhomogeneous partial shading \\
F5 & PV array mismatch & 15\% open circuit in PV array \\
F6 & MPPT/IPPT controller fault & $-20\%$ gain parameter of PI controller in MPPT/IPPT controller of the boost converter \\
F7 & Boost converter controller fault & $+20\%$ in time constant parameter of PI controller in MPPT/IPPT controller of the boost converter \\
\bottomrule
\end{tabular}
\end{table}
The collected measurements exhibit high-frequency noise, with additional disturbances and variations in temperature and insolation observed during and between the experiments. Moreover, the presence of MPPT/IPPT modes can adversely impact the detection of low-magnitude faults. Critical faults sometimes led to operational interruptions or system shutdowns, highlighting the challenge of detecting
faults before they escalate to complete failure. The variables collected in each scenario are presented in Table~\ref{tab:3}.
\begin{table}[h]
\centering
\caption{Descriptions of the variables under consideration.} 
\label{tab:3}
\begin{tabular}{lp{10cm}}
\toprule
\textbf{Variable} & \textbf{Description} \\
\midrule
Time & Time in seconds, average sampling    \( T_s = 0.9989 \) \textmu s \\
Ipv & PV array current measurement \\
Vpv & PV array voltage measurement \\
Vdc & DC voltage measurement \\
ia, ib, ic & 3-Phase current measurements \\
va, vb, vc & 3-Phase voltage measurements \\
Iabc & Current magnitude \\
If & Current frequency \\
Vabc & Voltage magnitude \\
Vf & Voltage frequency \\ 
\bottomrule
\end{tabular}
\end{table} 

In this study, four process variables that are sensitive to the system faults are considered, which are: PV current ($\text{Ipv}$), PV voltage ($\text{Vpv}$),
DC-link voltage ($\text{Vdc}$), and filter voltage ($\text{Vf}$), hereafter denoted
$v \in \{\text{Ipv},\text{Vpv}, \text{Vdc},\text{Vf}\}$.

\subsection{Methods}
\label{sec:3}

\subsection{Mathematical methods}
In this section, we present the theory underlying the mathematical methods used to estimate the kernel parameters without supervision for fault detection in a GCPVS. The kernel spaces and RKHS are described in Section~\ref{sec:method_RKHS}. Data pre-treatment Section~\ref{sec:pre-treatment}, benchmark methods for unsupervised fault detection in photovoltaic systems in Section~\ref{sec:other unsup models}. The proposed methodology and framework are presented in Section~\ref{sec:proposed approach} and Section~\ref{sec:proposed framework}, respectively. And finally, evaluation metrics in Section~\ref{sec:evaluation metrics}.
\subsubsection{Kernel methods}
\label{sec:method_RKHS}

The following definitions establish the theoretical foundations of the kernel-based monitoring framework proposed in Section \ref{sec:prop_method}. 

Given a dataset of process observations, each measurement can be represented as a point in a $d$-dimensional real-valued space, where $d$ corresponds to the number of monitored process variables. Let \( \mathcal{X} \subset \mathbb{R}^d \) denote this input space, and let 
\( \mathbf{x}, \mathbf{x}' \in \mathcal{X} \) represent any two such observation vectors drawn from it.
\newtheorem{definition}{Definition}
\newtheorem{theorem}{Theorem}
\begin{definition}
A \textbf{positive definite (p.d.) kernel} on a set \(\mathcal{X}\) is a function
\[
k : \mathcal{X} \times \mathcal{X} \to \mathbb{R}
\]
that is symmetric
\[
k(\mathbf{x}, \mathbf{x}') = k(\mathbf{x}', \mathbf{x}), \quad \forall \mathbf{x}, \mathbf{x}' \in \mathcal{X},
\]
and satisfies
\[
\sum_{i=1}^N \sum_{j=1}^N a_i a_j k(\mathbf{x}_i, \mathbf{x}_j) \geq 0
\]
for any \(N \in \mathbb{N}\), vectors
\(
\{\mathbf{x}_1, \ldots, \mathbf{x}_N\} \subset \mathcal{X},
\)
and coefficients \(a_1, \ldots, a_N \in \mathbb{R}\).
\end{definition}

\begin{theorem} 
Let \( \mathcal{X} \) be a nonempty set. A function  
\[k : \mathcal{X} \times \mathcal{X} \to \mathbb{R}\]  
is a positive definite kernel if and only if there exists a Hilbert space \( \mathcal{H} \) and a mapping  
\[\phi : \mathcal{X} \to \mathcal{H}\]  
such that
\begin{equation}
    k(\mathbf{x}, \mathbf{x}') = \langle \phi(\mathbf{x}), \phi(\mathbf{x}') \rangle_{\mathcal{H}}, \quad \forall \mathbf{x}, \mathbf{x}' \in \mathcal{X}.
\end{equation}
\end{theorem}
For more details of the given concepts, we refer readers to \cite{aronszajn1950theory}.
\begin{definition}
Let \( \mathcal{H} \) be a Hilbert space of functions on \( \mathcal{X} \).  
A kernel \( k \) is called a \textbf{reproducing kernel} of \( \mathcal{H} \) if  

\[k(\cdot, \mathbf{x}) \in \mathcal{H}, \quad \forall \mathbf{x} \in \mathcal{X},\]
and for any \( f \in \mathcal{H} \),  

\[f(\mathbf{x}) = \langle f, k(\cdot, \mathbf{x}) \rangle_{\mathcal{H}}.\]

If such a kernel exists, \( \mathcal{H} \) is called a \textbf{reproducing kernel Hilbert space (RKHS)}.
\end{definition}
\begin{definition}
Consider a positive definite kernel \( k(\mathbf{x}, \mathbf{x}') \) associated with an RKHS \( \mathcal{H} \).
The \textbf{kernel trick} implicitly maps the data vectors
\[\mathbf{x} \in \mathcal{X}\]
to a high-dimensional feature space via
\[\phi : \mathcal{X} \to \mathcal{H}.\]
Linear models can then be constructed in \( \mathcal{H} \) as
\[f(\mathbf{x}) = \langle \mathbf{w}, \phi(\mathbf{x}) \rangle_{\mathcal{H}}.\]
\end{definition}
Where $\mathbf{w} \in \mathcal{H}$ is the weight vector that parameterizes the linear model in the feature space $\mathcal{H}$, and $\langle \cdot,\cdot \rangle_{\mathcal{H}}$ denotes the inner product in $\mathcal{H}$.

\subsubsection{Kernel functions utilized in the study}
The central computational object in all kernel methods is the kernel matrix. Given a set of samples $\{\mathbf{x}_1, \ldots, \mathbf{x}_n\} \subset \mathcal{X}$, the kernel matrix $\mathbf{K} \in \mathbb{R}^{n \times n}$ has entries $K_{ij} = k(\mathbf{x}_i, \mathbf{x}_j)$, encoding the pairwise similarity between all samples in the dataset. This matrix serves as the basis for both kernel parameter optimization and the monitoring statistic introduced in Section~\ref{sec:proposed approach}. The specific kernel functions used to populate $\mathbf{K}$ in this study are defined below.

In our proposed framework, pairwise similarities between feature vectors \( \mathbf{x}, \mathbf{x}' \in \mathbb{R}^d \) are computed using kernel functions \(
k(\mathbf{x}, \mathbf{x}'),
\) which define the entries in the kernel matrix \(
\mathbf{K} \in \mathbb{R}^{n^{(w)} \times n^{(w)}}\)as \(
K_{ij} = k(\mathbf{x}_i, \mathbf{x}_j).
\) To evaluate the influence of different similarity structures on fault detection, multiple kernel functions are considered. A flexible class of covariance functions is given by the Mat{\'e}rn kernel, defined as:
\begin{equation}
    k_M(\mathbf{x}, \mathbf{x}'; \boldsymbol{\theta}) =  \gamma^2 \frac{2^{1-\nu}}{\Gamma(\nu)} \left( \sqrt{2\nu}\frac{\Vert\mathbf{x} - \mathbf{x}'\Vert}{\sigma} \right)^{\nu} K_{\nu} \left( \sqrt{2\nu}\frac{\Vert\mathbf{x} - \mathbf{x}'\Vert}{\sigma} \right) ,
\end{equation}
where $\boldsymbol{\theta}$ is the hyperparameter vector and includes $\{ \gamma^2, \sigma, \nu \}$, denoting amplitude, length scale and smoothness parameter respectively, $K_{\nu}$ is the modified Bessel function of the second kind and $\Gamma()$ is the Gamma function \cite{porcu2024matern}.  When $\nu$ is $\{ \frac{1}{2}, \frac{3}{2}, \frac{5}{2}, ...\}$, the Mat{\'e}rn kernel takes simpler forms that are easier to evaluate. If $\nu$ has a smaller value (i.e. $\frac{1}{2}$), the kernel is very rough. As $\nu$ approaches $\infty$, the smoother Gaussian kernel is obtained
\begin{equation}
    k_G(\mathbf{x}, \mathbf{x}'; \boldsymbol{\theta}) = \gamma^2 \exp \left( - \frac{\Vert\mathbf{x} - \mathbf{x}'\Vert^2}{2\sigma^2}\right) \text{.}\label{eq:Gauss_kernel}
\end{equation}
Here $\gamma$ is the kernel amplitude parameter and is related to the length scale parameter $\sigma$ by $\gamma=\frac{1}{2\sigma^2}$. Also known as the Radial Basis Function (RBF) kernel, it is the most commonly used kernel and is employed in this study. This kernel captures nonlinear similarities by measuring the exponential decay of the Euclidean distance between samples. 

Meanwhile, the heavier-tailed behavior can be obtained by the Cauchy kernel
\begin{equation}\label{eq:cauchy}
 k_C(\mathbf{x}, \mathbf{x}'; \boldsymbol{\theta}) =
\gamma^2 \left(1 + \frac{\Vert \mathbf{x}-\mathbf{x}'\Vert^2}{\sigma^2}\right)^{-1}.
\end{equation}
The resulting kernel matrix $\mathbf{K} \in \mathbb{R}^{n^{(w)} \times n^{(w)}}$, where $n^{(w)}$ is the window size, serves as a similarity graph for the data in the feature space. 

\subsection{Data pretreatment}
\label{sec:pre-treatment}
Before modeling, raw process signals often require pre-treatment to enhance feature extraction and improve model robustness. In this study, we employ frequency transformation techniques, such as the Fourier Transform (FT), to convert time-domain signals into the frequency domain \cite{oppenheim1999discrete}. FT features such as magnitude, frequency, variance, and residual statistics can be extracted over short time frames, enabling window-based monitoring with improved computational performance. This transformation is particularly effective for isolating cyclic patterns, vibrations, and other frequency-specific anomalies that may be obscured in the original signal, providing a more informative input for the subsequent kernel-based monitoring model. 

In the present study, the transformed signals are the features of the healthy and faulty variables being evaluated. When the continuous signal is replaced with a set of samples, a discrete Fourier transform (DFT) is applied \cite{mark2012feature}. DFT is defined as:
\begin{equation}   \text{FT}_{\textrm{k}}=\sum_{n=0}^{N-1}x_ne^{-i2\pi\frac{k}{N}n},
\label{eq:DFT_equation}
\end{equation}
where $\text{FT}_{\textrm{k}}$ represents the transformed sequence, $N $ is the sequence length, $x:=(x_0,\dots,x_{N-1})^T$ is the original sequence, and $ \frac{k}{N}$ is the frequency \cite{jorry2025statistical}. 

To capture the spectral characteristics of photovoltaic system dynamics, a frequency-domain feature-extraction approach based on discrete Fourier decomposition is adopted. For each signal and time window, the observed signal is decomposed into dominant Fourier components and a residual term as follows:
\begin{equation}
    x_t = \sum^{I}_{i=0} \text{FT}_{i,t} + r_t,
    \label{eq:DFT_decomp}
\end{equation}
where $\text{FT}_{i,t}$ denotes the $i^{\text{th}}$ dominant Fourier component within window $t$, and $r_t$ represents the residual signal capturing non-periodic variations.
The window length defines the number of samples per batch and determines the temporal resolution of the decomposition.

For each window $t$, a set of statistical descriptors of the dominant Fourier components and the residual signal is used to form a feature vector. These features were first introduced in \cite{jorry2025statistical}, specifically for each selected process variable $v \in \{\text{Ipv},\text{Vpv}, \text{Vdc},\text{Vf}\}$, the Fourier transform is computed within each window $t$. Let $\text{FT}_{i,t}^{v})$ denote the $i$-th retained Fourier component of variable $v$. Three statistical descriptors are extracted from each retained component:
\[
 \text{M}_{i,t}^{v} =\text{Max}(\text{FT}_{i,t}^{v}),
\]
\[
 \text{F}_{i,t}^{v} = \text{Freq}(\text{FT}_{i,t}^{v}),
 \]
 \[
\text{V}_{i,t}^{v} = \text{Var}(\text{FT}_{i,t}^{v}),
 \]
where $\text{M}_{i,t}^{v},\text{F}_{i,t}^{v},$ and $\text{V}_{i,t}^{v}$ denote the maximum magnitude, dominant frequency, and variance of the $i$-th Fourier component, respectively. In addition, the standard deviation of the residual signal $\sigma(r_t^{v})$ is also included. The resulting feature vector is then:
\begin{equation}
 \mathbf{x}_t = \left[\text{M}_{1,t}^{v},\text{F}_{1,t}^{v},\text{V}_{1,t}^{v}\ \ldots,\text{M}_{I,t}^{v},\text{F}_{I,t}^{v},\text{V}_{I,t}^{v}, \sigma(r_t^{v})\right].
 \label{eq:FFT_features}
\end{equation}
For further description of the feature notation, refer to the Table \ref{tab:feature_notation}.

\begin{table}[H]
\centering
\caption{Feature notation used throughout the study. Superscripts 
denote the process variable; subscripts denote the Fourier 
component index and window $t$.}
\begin{tabular}{lll}
\hline
\textbf{Notation} & \textbf{Feature} & \textbf{Description} \\
\hline
$\text{M}_{i,t}^{v}$ & Maximum magnitude & 
    Peak amplitude of the $i$-th Fourier component \\
    & & of variable $v$ in window $t$ \\
$\text{F}_{i,t}^{v}$ & Dominant frequency & 
    Frequency at peak amplitude \\
$\text{V}_{i,t}^{v}$ & Variance & 
    Spread of the $i$-th Fourier component \\
$\sigma(r_t^{v})$ & Residual standard deviation & 
    Standard deviation of the residual signal \\
\hline
\multicolumn{3}{l}{$v \in \{\text{Ipv},\text{Vpv},\text{Vdc},\text{Vf}\}$; 
    $i = 1, 2, \ldots, I$} \\
\hline
\end{tabular}
\label{tab:feature_notation}
\end{table}

\subsection{Other unsupervised fault detection methods}
\label{sec:other unsup models}
In this section, we highlight the traditional baseline and state-of-the-art methods used for comparison with the proposed method.

\subsubsection{One Class Support Vector Machine}
\label{subsec:oc-svm}
The OCSVM is an unsupervised machine learning one-class classifier and a special case of multi-class SVM \cite{harrou2019unsupervised}. It works by establishing a "maximum-margin hyperplane" based on normal training data and classifies new samples as normal or faulty depending on their distance from a given boundary \cite{scholkopf2001estimating}. The OCSVM uses the hyperplane as the decision boundary, which is equidistant to the closest inliers and outliers \cite{scholkopf2018learning}. Distinguishing new observations as normal or anomalous is dependent on their location in relation to a hyperplane. This involves solving an optimization problem defined by:
\begin{equation}
    \min_{\boldsymbol{\omega}\gamma\rho} \left( \frac{1}{2} \boldsymbol{\omega}^T \boldsymbol{\omega} - \rho + \frac{1}{\nu l} \sum_{i=1}^{l} \gamma_i \right)  
    \label{eq:ocsvm_objective}
\end{equation}
subject to:
\begin{equation}
    \boldsymbol{\omega}. \; \Psi(\mathbf{x}) > \rho - \gamma
    \label{eq:ocsvm_constraints}
\end{equation}
where $\boldsymbol{\omega}$ is a weight vector, $\nu$ a regularization parameter,$\gamma$ a slack variable, $l$ is the number of observations, and $\rho$ an offset term. The regularization parameter prevents overfitting, the slack variable manages deviations in observations, and the offset term determines the distance between the origin and mapped samples. The OCSVM decision function
, denoted $\mathcal{F}$, assigns a value of $-1$ to anomalies and a value of $1$
to typical data points based on their position relative to the hyperplane \cite{harrou2019unsupervised}
\begin{equation}
    \mathcal{F}(\mathbf{x}) = \text{sign} \left(\boldsymbol {\omega} \cdot \Psi(\mathbf{x}) - \rho \right).
    \label{eq:ocsvm_eq1}
\end{equation}

The OCSVM transforms the original data into a higher-dimensional feature space using the function $\Psi$. Within this space, the hyperplane is defined by the expression $\frac{\rho}{\|\boldsymbol{\omega}\|^2}$, denoting the Euclidean distance between the origin and the support vector point. The goal is to maximize this expression. For precise classification, the OCSVM addresses/resolves a quadratic optimization problem, as presented in Eq.\eqref{eq:ocsvm_eq1}, to uncover the optimal hyperplane that efficiently distinguishes between the data points
\begin{equation}
\min_{\boldsymbol{\omega}, \boldsymbol{\gamma}, \rho}
\left( \frac{\|\boldsymbol{\omega}\|^2}{2} - \rho + \frac{1}{\nu l} \sum_{i=1}^{l} \gamma_i \right).
\end{equation}

The OCSVM algorithm aims to maximize the margin of separation between the origin and the transformed data samples in the feature space. The approach involves optimizing the expression $\|\boldsymbol{\omega}\|^{2}/2-\rho$ while minimizing the average of slack variables $\boldsymbol{\gamma}$. Utilizing the Gaussian kernel (also known as the radial basis function kernel), the algorithm can capture complex relationships among data points, thereby improving the distinction between faulty and normal observations in the feature space.

\subsubsection{Isolation Forest}
\label{subsec:oc-IForest}

The Isolation Forest (iForest) is an OC classifier presented in \cite{liu2008isolation}, that efficiently handles a variety of data types. It forms a collection of decision trees, with each tree trained on a randomly chosen subset of the data \cite{harrou2026semi}. Anomalies are identified based on the sample's degree of normality, measured by the number of splits needed to isolate it in the trees \cite{harrou2024automatic}. Anomalous samples tend to have shorter average path lengths to the tree roots, as they are isolated by fewer splits than normal observations. The related anomaly score $S$ can then be evaluated for a given sample $\mathbf{x}$ from a dataset of $N$ samples. The iForest algorithm has been successfully applied to various anomaly detection tasks and has shown promising results \cite{buschjager2022randomized,dairi2022efficient,mckinnon2020comparison}. The anomaly score $S$ denoted as $S$ for a sample $\mathbf{x}$ drawn from a dataset of $N$ observations is defined as:
\begin{equation}
    S(\mathbf{x},N) = 2^{-\frac{E\left[h(\mathbf{x})\right]}{C(N)}},
    \label{eq:iForest_eq1} 
\end{equation}
where $h(x)$ represents the path length from the leaf that includes the sample $x$ to its root. The term $E\left(h\left(x\right)\right)$ denotes the average path length for all the isolation trees that form the iForest. The anomaly score is determined based on the inverse of this average path length, with a larger score indicating a higher likelihood of being an anomaly. The normalization and comparison average $E\left(h\left(x\right)\right)$ is calculated using the following formula:
\begin{equation}
    C\left(N\right) = 2H\left(N-1\right) - \frac{2\left(N-1\right)}{N}.
    \label{eq:iForest_eq2}
\end{equation}
Here, \( C(N) \) serves as a normalization factor in the anomaly score calculation to account for the dataset size. Here, \( H(N - 1) \) represents the harmonic number, which can be estimated as \(
H(N - 1) \approx \ln(N - 1) + 0.5772156649 \).
Generally, the IForest algorithm provides anomaly scores, where higher scores indicate a higher likelihood of a sample being an anomaly. Despite its computational efficiency and flexibility across data distributions, iForest operates on individual sample path lengths in the original feature space and does not explicitly model the joint nonlinear structure of the process --- a limitation that motivates the kernel-based approach proposed in Section {\ref{sec:proposed approach}}.

\subsubsection{Local Outlier Factor (LOF)}
\label{subsec:lof}

The Local Outlier Factor (LOF) is an OC classifier presented in \cite{breunig2000lof}, which is based on the local density deviation of the observed point to its k-nearest neighbors. It calculates the ratio of a point's local density to the density of its neighbors, flagging data points with a significantly lower density as outliers. The underlying assumption is that points similar to their neighbors are normal, while dissimilar points are outliers. The LOF of a sample \( \mathbf{x} \) can be computed using the following formula:
\begin{equation}
    S(\mathbf{x}) = \text{LOF}_k(\mathbf{x}) = \frac{1}{|N_k(\mathbf{x})|} \sum_{q \in N_k(\mathbf{x})} \frac{{lrd}_k(\mathbf{x}_q)}{{lrd}_k(\mathbf{x})}.
    \label{eq:LOF_eq1}
\end{equation}

The LOF algorithm determines the local density of a point by considering its k-distance neighborhood, denoted as \( N_k(\mathbf{x}) \). By comparing a point's density to its neighbors' densities, the LOF algorithm identifies points that are substantially less dense than their surroundings, indicating potential outliers. The calculation of LOF is defined as follows:
\begin{equation}
    N_k(\mathbf{x}) = \{\mathbf{ x}_q \in \mathcal{X} \mid d(\mathbf{x}, \mathbf{x}_q) \leq d_k(\mathbf{x}) \}.
    \label{Nk}
\end{equation}

The local reachability density ($lrd$), denoted as \({lrd}_k(\mathbf{x}) \), is a measure used in the LOF algorithm to quantify the isolation of a data point. It is calculated by taking the ratio of the average reachability distance of the point's k-neighborhood to the k-distance of the point. The $lrd$ is used to compute the LOF, which indicates the outlierness of a point. A higher LOF value signifies a higher degree of outlierness. Mathematically, the $lrd$ of a point \( x \) is defined as:
\begin{equation}
    {lrd}_k(\mathbf{x}) = \left( \frac{1}{|N_k(\mathbf{x})|} \sum_{q \in N_k(\mathbf{x})} {rd}_k(\mathbf{x}, \mathbf{x}_q) \right)^{-1}
    \label{eq: LOF_LRD}
\end{equation}
where \( N_k(\mathbf{x}) \) is the k-neighborhood of point \( \mathbf{x} \), and \( \text{RD}_k(\mathbf{x}, \mathbf{x}_q) \) is the reachability distance between point \(\mathbf{x} \) and its neighbor \(\mathbf{x}_q \).

The reachability distance \( {rd}_k(\mathbf{x},\mathbf{x}_q) \) is defined as the maximum value between the k-distance of the neighbor \(\mathbf{x}_q \) and the distance between point \(\mathbf{x} \) and its neighbor \( \mathbf{x}_q \) \cite{luan2021out}. An outlier is identified as a point with a higher LOF score than its neighbors, indicating a lower local density. The threshold for outlier identification relies on the distribution of LOF values across all dataset points.

\subsection{The proposed method for unsupervised fault detection}
\label{sec:prop_method}
Since labeled fault data are not readily available in real-life applications, we evaluate kernel parameters using unsupervised criteria that measure the quality of the data distribution in the feature space.
\label{sec:proposed approach}

\subsubsection{Variance maximization}
\label{ssec:varmax}
In the proposed variance-maximization approach, the data is first mapped into the RKHS,  \(\mathcal{H}\). Let $\mathbf{X} \in \mathbb{R}^{N\times d}$ be the multivariate time series with rows $\mathbf{x}_i^\top \in \mathbb{R}^d$ ($i=1,\ldots, N$), and sampling period $T_s$. We apply sliding windows of length $w$ (samples) and step $s$. Let's denote the window ending at sample $t$ as follows:
\begin{equation}
\mathbf{W}_t = \{\mathbf{x}_{t-w+1},\ldots,\mathbf{x}_t\}\in\mathbb{R}^{w\times d}.
\end{equation}

For a window $\mathbf{W}_t$, we compute the pairwise squared Euclidean distance matrix $\mathbf{D}\in\mathbb{R}^{w\times w}$,
\begin{equation}
D_{ij}=\|\mathbf{x}_{t-w+i}-\mathbf{x}_{t-w+j}\|_2^2 .
\end{equation}

The Gaussian (Eq.~\ref{eq:Gauss_kernel}), Mat{\'e}rn or Cauchy kernel  is then applied to the pairwise distance
matrix $\mathbf{D}$, giving kernel matrix entries:
\begin{equation}
 K_{ij}(\sigma) = k_G\!\left(\mathbf{x}_i, \mathbf{x}_j;\, \sigma\right),
\end{equation}
where $\sigma > 0$ is the length-scale parameter to be optimized
$[\mathbf{K}(\sigma)]_{ij} = k_G(\mathbf{x}_i, \mathbf{x}_j;\, \sigma)$.

We then center the kernel matrix $\mathbf{K}$ in feature space using the centering matrix $\mathbf{H}=\mathbf{I}_w - \tfrac{1}{w}\mathbf{1}\mathbf{1}^\top$:
\begin{equation}
\mathbf{K}_c(\sigma)=\mathbf{H}\,\mathbf{K}(\sigma)\,\mathbf{H}.
\end{equation}

The element-wise equivalent of this is given by:
\begin{equation}
\mathbf{K}_c = \mathbf{K} - \frac{1}{w}\mathbf{1}\mathbf{1}^\top \mathbf{K} - \frac{1}{w}\mathbf{K}\mathbf{1}\mathbf{1}^\top + \frac{1}{w^2}\mathbf{1}\mathbf{1}^\top \mathbf{K} \mathbf{1}\mathbf{1}^\top.
\end{equation}
For selecting $\sigma$ and for monitoring, the first is the top-$l$ eigenvalue sum that computes the eigenvalues $\lambda_1\ge\lambda_2\ge\cdots\ge\lambda_w$ of a symmetric $\mathbf{K}_c$. For a chosen $l$, it is computed by:
    \begin{equation}
S_l(\sigma)=\sum_{i=1}^{l}\lambda_i\big(\mathbf{K}_c(\sigma)\big).
    \label{eq:eigensum}
    \end{equation}
For monitoring, the kernel matrix is computed for each window with the initial $\sigma$. The monitoring statistic is the Frobenius norm:
    \begin{equation}
    M(\sigma)=\|\mathbf{K}_c(\sigma)\|_F=\sqrt{\sum_{i,j} \big(K_c(\sigma)_{ij}\big)^2}=\sqrt{\mathrm{trace}\big(\mathbf{K}_c(\sigma)^\top \mathbf{K}_c(\sigma)\big)}.
    \end{equation}
Hence, during optimization, we maximize $-\|\mathbf{K}_c\|_F$ when minimizing total kernel energy.

The logarithmic grid search $\Gamma=\{ \sigma_1,\ldots,\sigma_m\}$ is then applied to the scoring function and computes the optimal $\sigma$ ($\sigma_{\text{opt}}$):
\begin{equation}
\sigma_{\text{opt}}=\arg\max_{\sigma\in\Gamma} S_l(\sigma),
\end{equation}
In the initial (baseline) window, then fix $\sigma=\sigma_{\text{opt}}$ for all subsequent windows. For each sliding window ending at time point $t_k$, the monitoring statistic is computed as:
\begin{equation}
m_{t_k}=M(\sigma_{\text{opt}})=\|\mathbf{K}_c(\sigma_{\text{opt}})\|_F .
\end{equation}

The first $n_{\mathrm{init}}$ windows is used as a baseline (e.g., first $6\,\mathrm{s} / T_s$) for threshold computation. We compute baseline mean and standard deviation as:
\[
\mu=\frac{1}{n_{\mathrm{init}}}\sum_{k=1}^{n_{\mathrm{init}}} m_{t_k},\qquad
\hat{\sigma}=\sqrt{\frac{1}{n_{\mathrm{init}}-1}\sum_{k=1}^{n_{\mathrm{init}}}(m_{t_k}-\mu)^2}.
\]
The control limits are then computed as:
\begin{equation}
\begin{aligned}
    \text{Upper control limit (UCL)} &= \mu + \kappa\hat {\sigma}, \\
    \text{Lower control limit (LCL)} &= \mu - \kappa\hat{ \sigma}.
    \label{eq:prop_thres}
\end{aligned}
\end{equation}

The scalar $\kappa$ determines the width of the control band. Setting $\kappa=2$ corresponds to a 95\% confidence interval, whereas $\kappa=3$ corresponds to a 99.7\% confidence interval under the assumption of normality. This study uses  $\kappa=2$ throughout to improve the sensitivity of the proposed framework to early structural changes during IPPT system operation. A 95\% control limit represents a commonly adopted compromise in statistical process monitoring, providing sufficient sensitivity to detect incipient faults while maintaining an acceptable false alarm rate during normal operation \cite{kourti1995process}.

We then declare the first window index $k^*$ such that $m_{t_{k^*}}\not\in[\mathrm{LCL},\mathrm{UCL}]$ as the detected change point. In principle, this index can be converted to physical time as $t_{\mathrm{det}}=t_{k^*}\cdot T_s$, where $T_s \approx 9.999 \times 10^{-6}$~s is the sampling interval of the GPVS-Faults dataset (sampling frequency $\approx100$~kHz). In the multivariate results presented in Section~\ref{ssec:Realdata}, detection performance is reported in terms of sample index rather than physical time, as this representation is independent of dataset-specific sampling rates and facilitates direct comparison with benchmark methods. Conversion to physical time is straightforward for practitioners using the relation above.
This detailed approach is summarized in the Algorithm \ref{alg:kernel_change_detection} below.

\begin{algorithm}[H] 
\caption{Unsupervised Kernel-Based Change Detection}
\label{alg:kernel_change_detection}
\begin{algorithmic}[1]
\Require $\mathbf{X} \in \mathbb{R}^{N \times d}$, window length $w$, step $s$, 
         grid $\Gamma$, baseline duration (samples), $T_s$

\State Acquire real-time process measurements $\mathbf{x}_t \in \mathbb{R}^d$ and partition into sliding windows $\{W_{t_k}\}$ of length $w$ with step $s$. 
\State For each window $W_{t_k}$, apply the DFT (Eq.~\ref{eq:DFT_equation}) and extract features per Eq.~\ref{eq:FFT_features}
\State Identify the initial baseline window $W_{t_1}$ as the first $w$ samples of the normal operation data.
\For{each $\sigma \in \Gamma$}
    \State Build $\mathbf{K}(\sigma)$ on $W_{t_1}$, center 
           $\mathbf{K}_c(\sigma)$, compute score $S_l(\sigma)$
\EndFor
\State $\sigma_{\text{opt}} \leftarrow \arg\max_{\sigma \in \Gamma}\, S_l(\sigma)$
\State Compute baseline indices $k = 1, \ldots, n_{\text{init}}$
\State Compute $\mu,\, \hat{\sigma}$ from $\{m_{t_k}\}_{k \leq n_{\text{init}}}$ 
       and set UCL, LCL per Eq.~\ref{eq:prop_thres} 
\For{each window $W_{t_k}$}
    \State Build $\mathbf{K}(\sigma_{\text{opt}})$, center 
           $\mathbf{K}_c(\sigma_{\text{opt}})$
    \State $m_{t_k} \leftarrow \|\mathbf{K}_c(\sigma_{\text{opt}})\|_F$
\EndFor
\State Return detected window index $k^*$, monitoring statistics $\{m_{t_k}\}$; physical detection time obtainable as $t_{\text{det}}=k^* \cdot T_s$
\end{algorithmic}
\end{algorithm}

\subsubsection{Contribution analysis for fault diagnosis}

To identify the process variables most responsible for a 
detected fault, individual variable contributions are computed 
through univariate kernel analysis applied to the fault window 
$\mathbf{W}_{k^*}$ --- the same window of length $w$ at which 
the monitoring statistic $m_{t_{k^*}}$ first exceeded the 
control limits. For each process variable 
$v \in \{\text{Ipv},\, \text{Vpv},\, \text{Vdc},\, \text{Vf}\}$, a kernel matrix is constructed from the $w$ observations of the variable 
$v$ alone within $\mathbf{W}_{k^*}$:
\begin{equation}
    K^{(v)}_{kl} = k\!\left(x_k^{(v)},\, x_l^{(v)};\, 
    \sigma_\text{opt}\right), \quad 
    k,l = 1, \ldots, w
    \label{eq:pv_kernel}
\end{equation}

The per-variable kernel matrix $\mathbf{K}^{(v)} \in 
\mathbb{R}^{w \times w}$ is then centered using the same 
centering matrix $\mathbf{H} = \mathbf{I}_w - 
\frac{1}{w}\mathbf{1}\mathbf{1}^\top$ as in 
Section \ref{ssec:varmax}:
\begin{equation}
    \mathbf{K}_c^{(v)} = \mathbf{H}\,\mathbf{K}^{(v)}\,\mathbf{H}.
    \label{eq:pv_centered}
\end{equation}

The raw contribution score of variable $v$ is the Frobenius 
norm of its centered per-variable kernel matrix:
\begin{equation}
    C_v = \left\|\mathbf{K}_c^{(v)}\right\|_F,
    \label{eq:contribution_raw}
\end{equation}
and the normalized contribution, which sums to one across all variables, is:
\begin{equation}
    \hat{C}_v = \frac{C_v}{\displaystyle\sum_{v} C_v}.
    \label{eq:contribution_norm}
\end{equation}

\subsection{Proposed framework}
\label{sec:proposed framework}
\begin{figure}[H]
    \centering
    \includegraphics[width=0.99\linewidth]
    {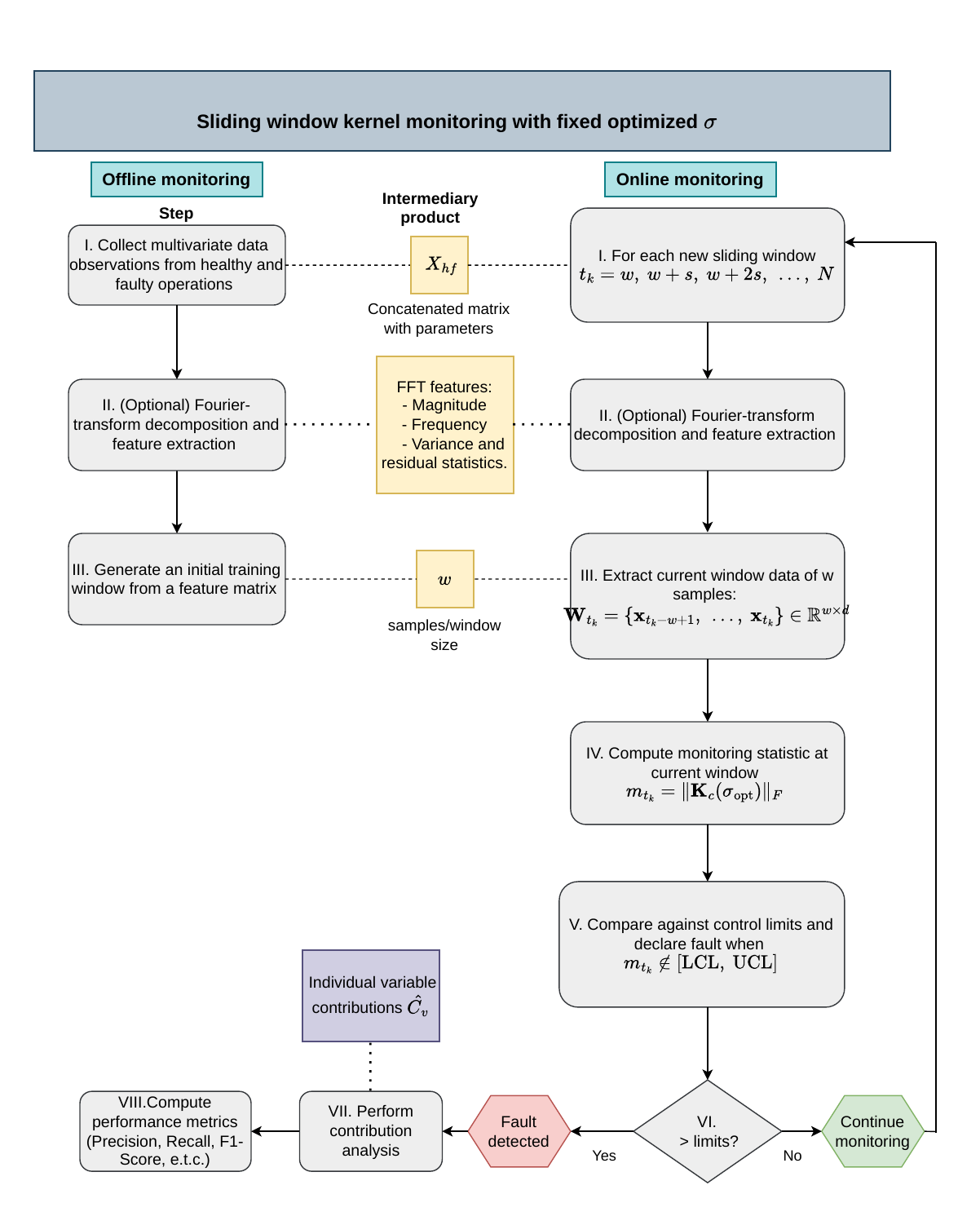}
    \caption{Proposed KNM fault detection workflow.}
    \label{fig:workflow}
\end{figure}

The diagram in Fig.\ref{fig:workflow} summarizes the proposed fault detection framework based on the mathematical methods presented in the preceding sections. The process data is collected under normal operating conditions in IPPT mode, where both univariate and multivariate process variables are considered. In cases where fault manifestation is primarily reflected as a change in signal periodicity rather than a mean shift, the raw time-domain signals are first transformed into the frequency domain via DFT decomposition, as described in Section {\ref{sec:pre-treatment}}. The resulting spectral features---maximum magnitude, dominant frequency, and variance of each Fourier component, together with residual standard deviation --- are assembled into a feature matrix $\mathbf{X}$.

The framework operates in two phases: an offline training phase and an online monitoring phase. During the offline phase, the first $w$ samples of the normal operation data from the initial training window using a selected kernel function --- either the RBF, Cauchy or Mat{\'e}rn kernel, and is subsequently centered in the feature space to ensure zero mean in the RKHS. The optimal length-scale parameter $(\sigma_{\text{opt}})$ is then determined unsupervisedly by evaluating candidate values over a logarithmic grid $\Gamma \subset[10^{-6},\,10^{6}]$, selecting the value that maximizes the sum of the top $l$ eigenvalues of the centered kernel matrix, as defined in Eq.(\ref{eq:eigensum}) the kernel feature space and requires no labeled fault data.

During the online monitoring phase,$(\sigma_{\text{opt}})$ is fixed, and the centered kernel matrix is recomputed for each new sliding window. The Frobenius norm of the centered kernel matrix serves as the monitoring statistic $m_{t_k}$, quantifying the total nonlinear variance of the current window in the kernel-induced feature space. At each time step, $m_{t_k}$ is compared against the upper and lower control limits derived from the baseline normal operation statistics, as defined in Eq.(\ref{eq:prop_thres}). When a value falls outside these limits, a fault is declared, and the detected change point is recorded. A contribution analysis is subsequently performed to identify the process variables most responsible for the detected fault, supporting fault diagnosis as described in Section {\ref{ssec:varmax}}.

\subsection{Evaluation of results}
\label{sec:evaluation metrics}
In this section, the ability of the proposed method to monitor PV system data is highlighted. The investigated scenarios are F$2$L (feedback sensor fault) and F$4$L (partial shading), where L stands for IPPT mode. In addition, the efficacy of the proposed approach is compared with other unsupervised learning methods, such as OCSVM \cite{harrou2019unsupervised}, iForest \cite{zhao2018hierarchical}, and LOF \cite{ding2018local}. They are trained using only fault-free data and are applied to cluster the test data into normal observations or anomalies. The parameter values used in the benchmark methods are listed in Table~\ref{tab:hyperparameters} in the Appendix and were determined using a grid search.   

The detection quality has been measured using the true positive rate (TPR), false positive rate (FPR), and accuracy, as in \cite{harrou2019unsupervised}. TPR (i.e., detection rate/recall) is the number of identified faults (i.e., true positives (TP)) over the total number of faults, P,
\begin{equation}
    \mathrm{Recall} = \frac{\mathrm{TP}}{\mathrm{TP} + \mathrm{FN}} = \frac{\mathrm{TP}}{\mathrm{P}},
\end{equation}
where FN represents false negatives or missed detections (Type II errors), the number of faults that do not exceed the threshold. FPR (also known as the false alarm rate, FAR) is the number of normal observations incorrectly detected as faults divided by the total number of fault-free data, N.
\begin{equation}
    \mathrm{FAR} = \frac{\mathrm{FP}}{\mathrm{FP} + \mathrm{TN}} = \frac{\mathrm{FP}}{\mathrm{N}},
\end{equation}
where FP is false positives or false alarms (i.e., Type I errors) and TN is true negatives. The detection accuracy, which represents the percentage of correctly detected faults, can be computed as:
\begin{equation}
    \mathrm{Accuracy} = \frac{\mathrm{TP}+\mathrm{TN}}{\mathrm{P} + \mathrm{N}}.
\end{equation}

The missed detection rate (MDR), also known as the false negative rate, is the probability that a true fault event goes undetected by a system, computed by:
\[
\mathrm{MDR}
= \frac{\mathrm{FN}}{\mathrm{TP} + \mathrm{FN}}.
\]
Apart from MDR, we also have a detection delay metric, which measures how long the detector waits after the true fault onset before raising an alarm, that is, \(
\text{detection delay} = k^* - k_f
\). Where $k^*$ is the first window where the monitoring statistic exceeds the control limits, and $k_f$ is the fault injection window index. A smaller delay means faster fault recognition, which is especially important for time-critical monitoring. MDR and detection delay have been used for the performance evaluation of a simulated experiment presented in Section \ref{ssec:Simres}.
And we have the detection precision, which is computed as follows:
\begin{equation}
    \mathrm{Precision} = \frac{\mathrm{TP}}{\mathrm{TP} + \mathrm{FP}}.
\end{equation}

The F1-score, a harmonic mean of precision and recall, provides a single metric to evaluate
the trade-off between the two metrics is given by:
\begin{equation}
\text{F1-Score} = 2 \times \frac{\text{Precision} \times \text{Recall}}{\text{Precision} + \text{Recall}}
\end{equation}

The Area Under the Curve (AUC) is a performance metric utilized to quantify detection performance. Generally, an AUC of 1 corresponds to a perfect detector, an AUC of 0.5 
corresponds to a random detector (no better than chance), and an AUC $> 0.9$ corresponds to good detection performance. The AUC is defined as:

\begin{equation}
\mathrm{AUC} = \frac{\mathrm{TPR} + (1 - \mathrm{FPR})}{2}
\label{eq:auc}
\end{equation}

\section{Results and discussion}
\label{sec:results}
In this section, we present the results of the two case studies that validate the effectiveness of the proposed method on simulated and real solar system datasets. In Section~\ref{ssec:Simres}, we show the results of the proposed framework on simulated data as described above, followed by the results of the IPPT switching mode real data in Section~\ref{ssec:Realdata}.

\subsection{Demonstration of proposed method: simulated data}
\label{ssec:Simres}
\subsubsection{Unsupervised multivariate fault detection}

\begin{figure}[H]
    \centering
    \includegraphics[width=0.70\linewidth]{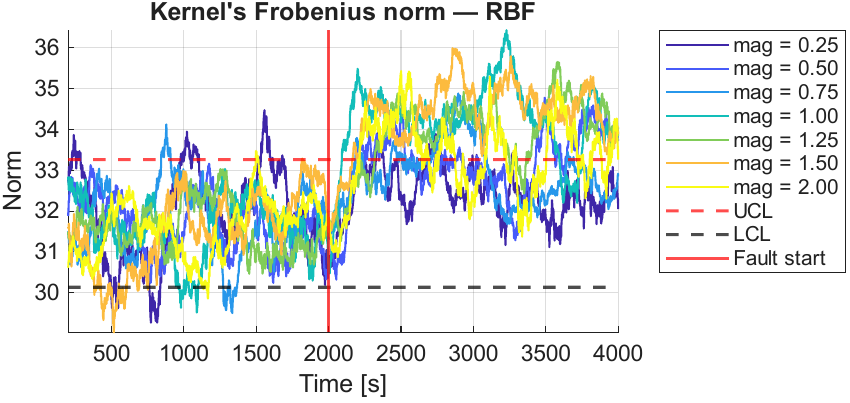}
    \caption{RBF kernel norm trajectories for simulated fault magnitudes (mag=0.25, 0.5, 0.75, 1, 1.25, 1.5, 2). Each colored line shows the Frobenius KNM over time for each fault magnitude using an RBF kernel, and the dashed horizontal lines mark the fixed healthy control limits computed from the pre-fault baseline.}
        \label{fig:simg_rbf}
\end{figure}

Fig.~\ref{fig:simg_rbf} displays the evolution of the KNM statistic under progressively increasing fault magnitudes for the RBF kernel. The plot shows that the separation between the healthy and faulty operating conditions becomes significant as the fault magnitude increases. At small fault magnitudes, the kernel norm trajectories remain close to the healthy baseline, indicating that subtle changes are introduced into the process dynamics. As the fault magnitude increases, the monitoring statistic exhibits larger deviations from its nominal behavior, resulting in clearer departures from the control limits and improved detectability.

The results verify that the kernel norm is sensitive to changes in the underlying data distribution. Because the monitoring statistic is computed from the centered kernel matrix, larger faults induce greater perturbations in the nonlinear similarity structure of the observations, which in turn produces larger changes in the kernel norm. This behavior is desirable from a fault-detection perspective because it establishes a monotonic relationship between fault severity and the monitoring response. Fig.~\ref{fig:kernels_case3} shows the results of the same experiment for Cauchy and Mat{\'e}rn kernels. The differences among the kernel functions are minimal. The RBF kernel produced the largest separation between the normal and faulty operating conditions, reflecting its strong sensitivity to local changes in sample similarity.

In contrast, the Mat{\'e}rn kernel exhibited a smoother response, providing a compromise between sensitivity and robustness to noise. While the Cauchy kernel showed a more gradual transition following fault occurrence, suggesting increased robustness to outliers but potentially reduced sensitivity to small fault magnitudes. These observations indicate that kernel selection influences the balance between early fault detection and resistance to misleading fluctuations.

\begin{figure}[H]
    \centering
        \includegraphics[width=0.75\linewidth]{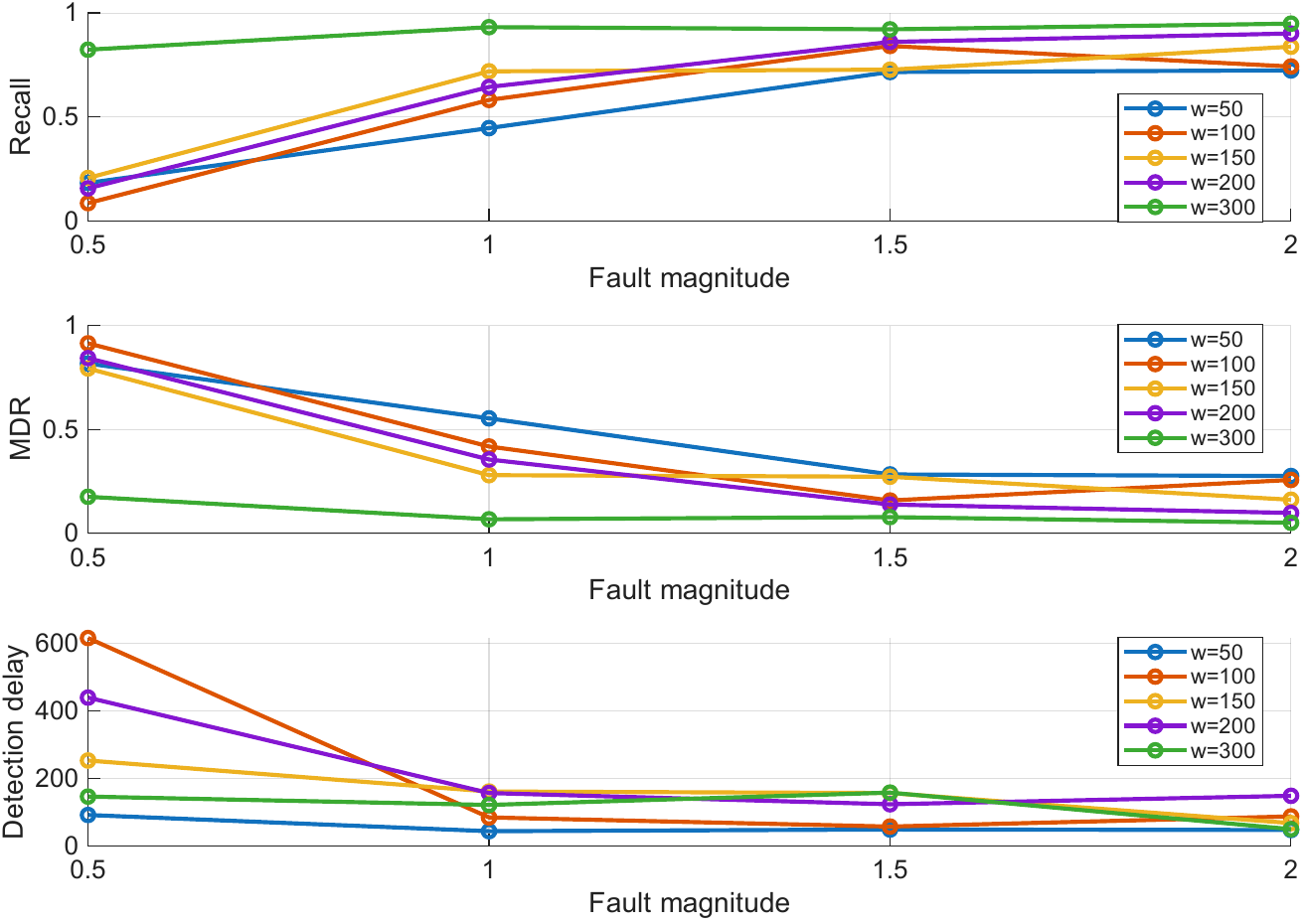}%
    \caption{Sensitivity performance for the RBF kernel as fault magnitude increases. Each plot shows one window size, with plotted metrics of mean recall, mean missed detection rate (MDR), and mean detection delay. Thresholds were fixed from the normal baseline period, so the plots compare how detection quality and timeliness evolve for different window lengths using the same kernel.}
    \label{fig:kernels_comp}
\end{figure}

Furthermore, Fig.~\ref{fig:kernels_comp} summarizes the performance metrics obtained for different sliding-window lengths ($w=50, 100, 150, 200,$ and $300$). Across all kernel types, recall generally increases with increasing fault magnitude, while the MDR decreases correspondingly. This trend confirms that larger faults are easier to identify because they produce stronger deviations from the healthy operating regime. The results further reveal a trade-off associated with window length selection. Smaller windows provide faster responsiveness because fewer observations are required before a fault-induced change influences the monitoring statistic. Consequently, smaller windows tend to yield lower detection delays. Thus, for a smaller $\delta_f$, a large window is needed to detect the fault, but when $\delta_f$ is large, the choice of window size becomes insignificant. 

Overall, intermediate window lengths ($w=100-200$) provided a favorable balance between detection speed and robustness. These windows were sufficiently large to suppress random fluctuations while remaining responsive to fault-induced changes. The findings, therefore, support the importance of performing window-length sensitivity analysis when designing kernel-based monitoring systems.

\subsection{Univariate and multivariate unsupervised KNM fault detection results}
\label{ssec:Realdata}
\begin{figure}[H]
    \centering
    \subfloat[Individual Ipv variable]{
    \label{fig:F2L_Ipv}%
        \includegraphics[width=0.52\linewidth]{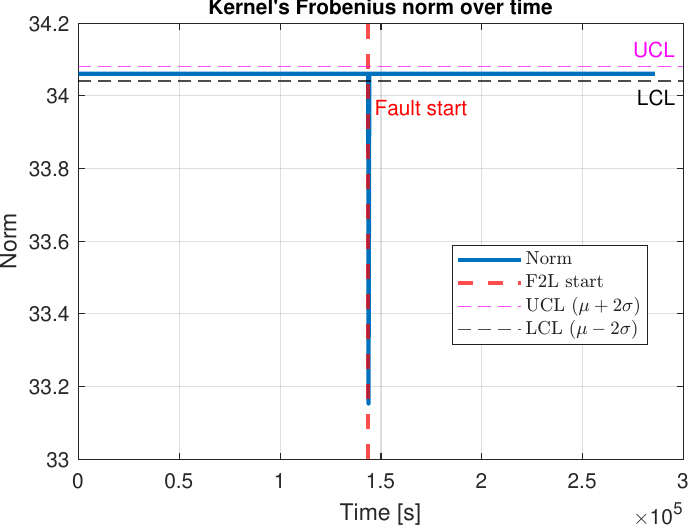}%
    }
    \subfloat[Combined 13 variables]{\label{fig:F2L_13}%
        \includegraphics[width=0.48\linewidth]{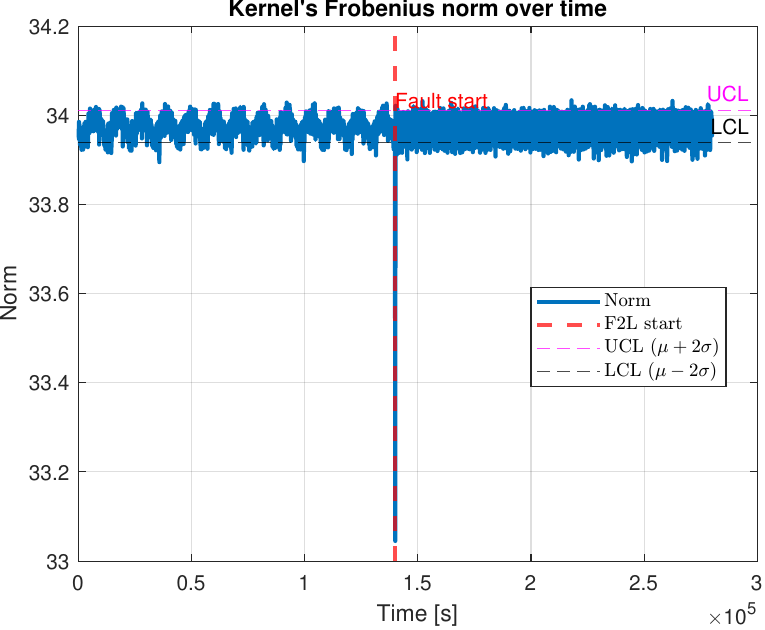}%
    }
    \hfill
    \subfloat[Individual Ipv variable]{\label{fig:F4L_Ipv}%
        \includegraphics[width=0.53\linewidth]{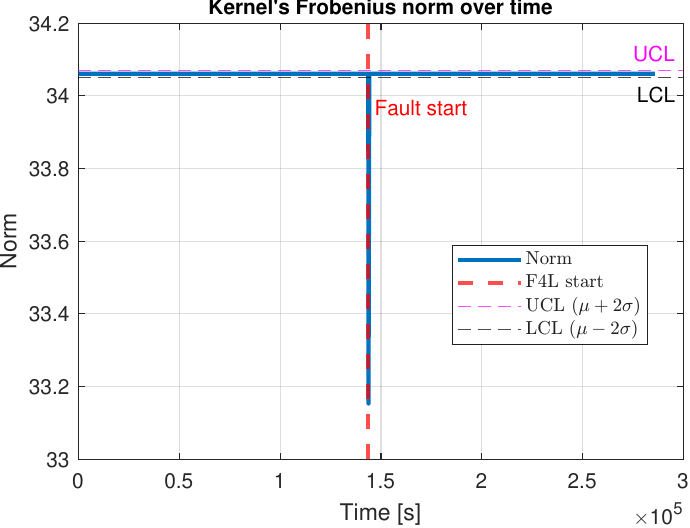}%
    }
    \subfloat[Combined 13 variables]{\label{fig:F4L_13}%
        \includegraphics[width=0.49\linewidth]{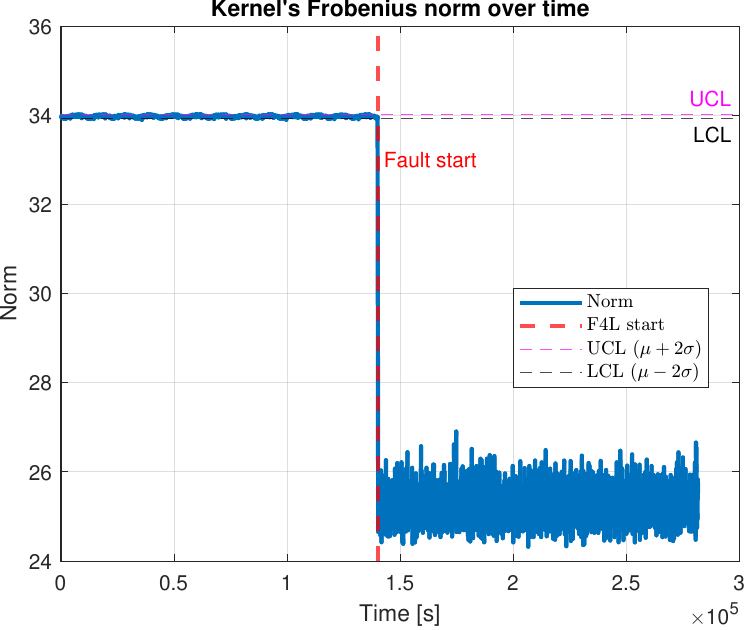}%
    }
    \caption{The Frobenius KNM for individual (Ipv) and combined (13 raw variables) signals under sensor (F2L) and partial shading (F4L) faults. A vertical red dashed line marks fault onset. (a) Ipv under F2L: the norm returns to baseline after fault onset. (b) Combined variables under F2L: clear separation between pre‑ and post‑fault behavior. (c) Ipv under F4L: similar return to baseline pattern. (d) Combined variables under F4L: an abrupt shift indicates fault detection.}
    \label{fig:method 3}
\end{figure}

Given the limitations of existing methods outlined in the introduction, we proposed a more advanced kernel-based approach in Section \ref{sec:proposed approach} to identify sensor and partial-shading faults. These are among the most prevalent faults in a solar PV system, and the same approach can be extended to other fault scenarios when needed. Fig.~\ref{fig:method 3} shows the results of this approach for the unaltered Ipv variable and all other 13 variables for F$2$L and  F$4$L faults. The results indicate that two faults are detected when the Ipv feature is used, as shown in Figs.~\ref{fig:F2L_Ipv} and \ref{fig:F4L_Ipv}. That is, when the fault occurs, the signal falls outside the control limits calculated using Eq.\eqref{eq:prop_thres} and then returns to its previous normal condition. This was also tested for other features, and the results were the same. The approach was also tested on all 13 combined raw variables, and the fault deviations could not be captured as shown in Fig.~\ref{fig:F2L_13}, but a clear abrupt drift could be found for F$4$L as in Fig.~\ref{fig:F4L_13}. From this, we decided to extract FFT features from four highly sensitive variables (Ipv, Vf, Vdc, and Vpv) based on previous literature \cite{chokr2023feature}, by decomposing each signal into FFT components as described in Section \ref{sec:pre-treatment}. 

Afterward, we apply a similar approach described in Section \ref{sec:proposed approach}, using the equations highlighted therein to identify the given faults using the new dataset after feature transformations. Fig.~\ref{fig:F2L_FFT3} below shows the results of the proposed KNM framework on the F$2$L fault by monitoring the Frobenius norm of the extracted FFT features for each selected variable. The results indicate that the fault is captured across all variables with minimal false alarms. In all cases, we observe that deviations remain stable during healthy operation, but suddenly deviate sharply upward or downward when the fault is introduced and cross the control limits. Among all variables, it can be seen that the fault is better captured in Figs.~\ref{fig:F2L_Ipv_FFT3} and \ref{fig:F2L_Vdc_FFT3}, indicating that these variables are susceptible to F$2$L. After the fault is captured, we can observe that the norm stabilizes in a particular state. Thus, a comparison can be made between the proposed method and traditional multivariate approaches, such as PCA, and its chart statistics T$^2$ and Q, which cannot separate this fault from healthy data \cite{bakdi2021real}.
\begin{figure}[H]
	\centering
	\subfloat[Ipv FFT features]{
    \label{fig:F2L_Ipv_FFT3}
		\includegraphics[width=0.50\linewidth]{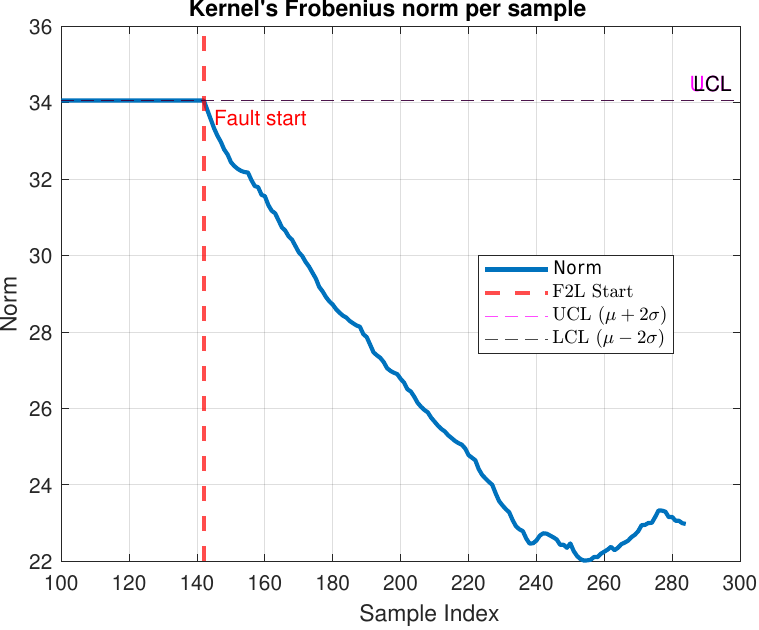}
	}
	\subfloat[Vf FFT features]{
		\includegraphics[width=0.50\linewidth]{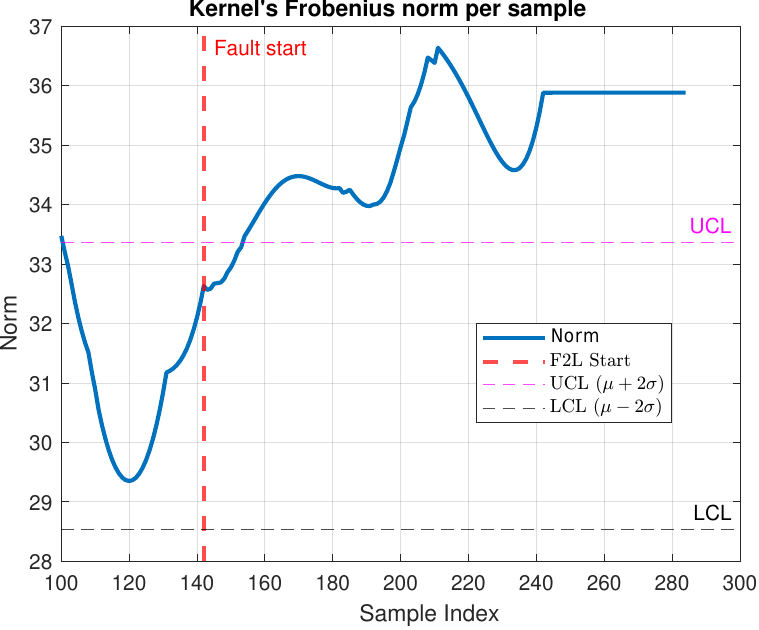}
	}\
    \subfloat[Vdc FFT features]{
    \label{fig:F2L_Vdc_FFT3}
		\includegraphics[width=0.50\linewidth]{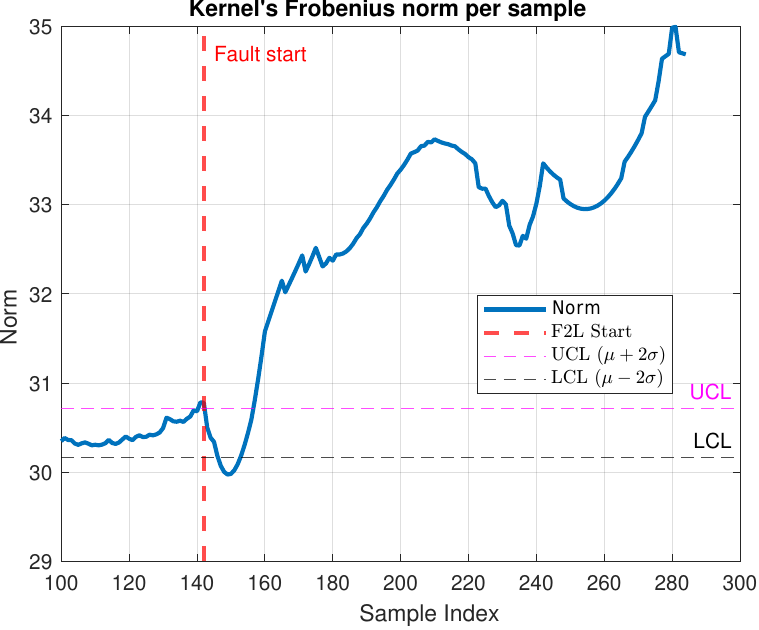}
	}
     \subfloat[Vpv FFT features]{
		\includegraphics[width=0.50\linewidth]{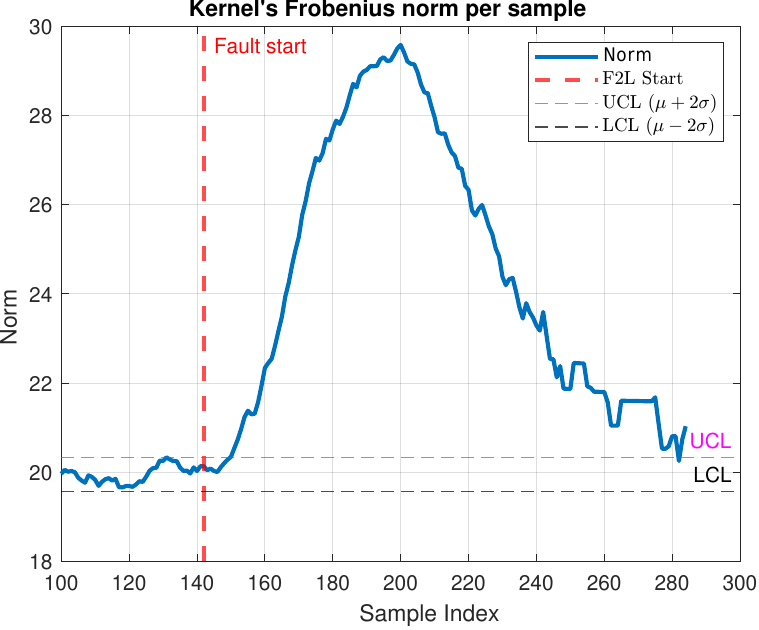}
    }
\caption{ The Frobenius KNM of FFT features for individual variables under the F2L fault scenario. The blue solid line shows the norm over sliding windows (samples 100–300), with fault onset at sample 141 (vertical red dashed line). Control limits (upper: magenta, lower: black) are derived from healthy operation. Each variable exhibits a distinct fault response: (a) Ipv drops below the LCL; (b) Vf rises above the UCL with a non‑monotonic peak; (c) Vdc shows sustained exceedance; (d) Vpv peaks sharply then returns toward the limits.}
\label{fig:F2L_FFT3}
\end{figure}

\begin{figure}[H]
	\centering
	\subfloat[Ipv Fault]{
    \label{fig:F4L_Ipv_FFT3}
		\includegraphics[width=0.50\linewidth]{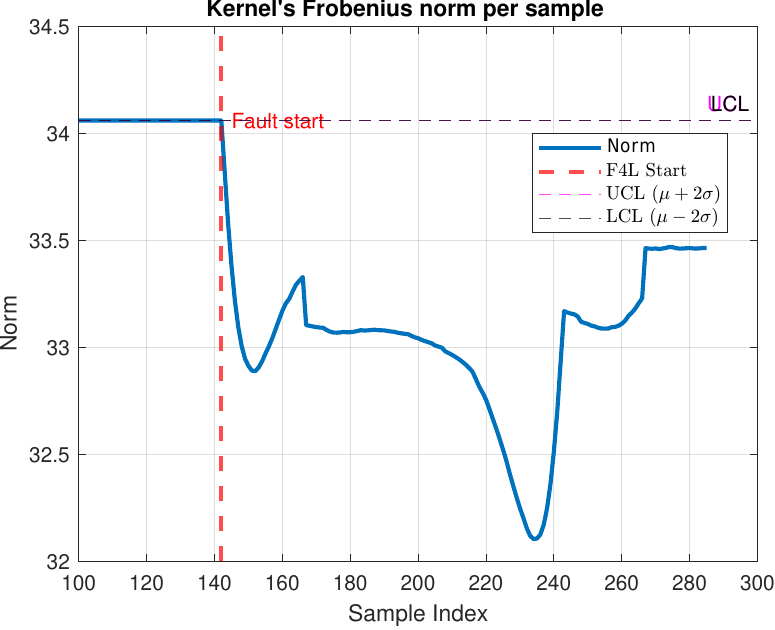}
	}
	\subfloat[Vf Fault]{
		\includegraphics[width=0.49\linewidth]{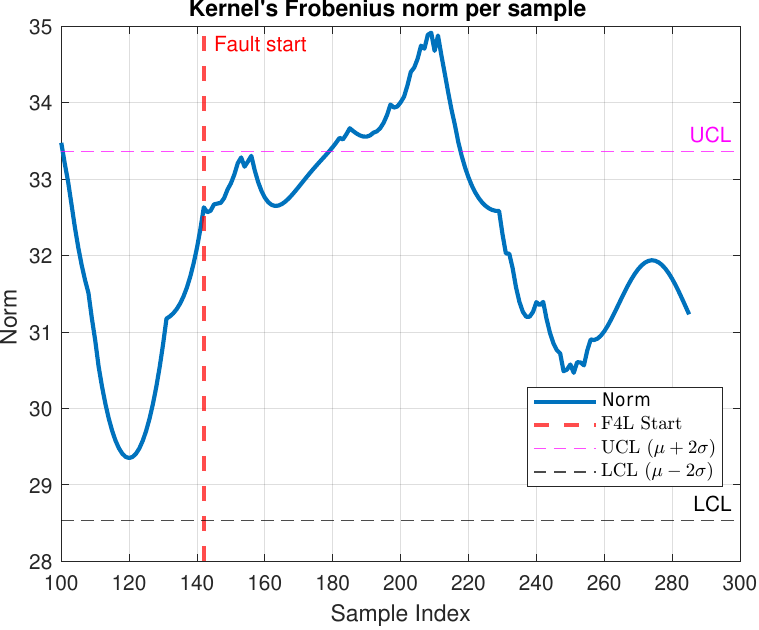}
	}\
    \subfloat[Vdc Fault]{
    \label{fig:F4L_Vdc_FFT3}
		\includegraphics[width=0.49\linewidth]{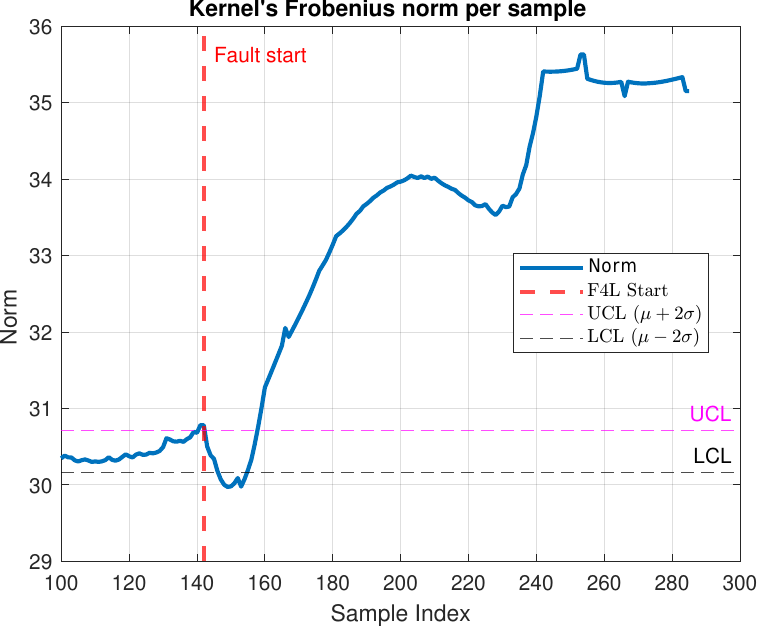}
	}
     \subfloat[Vpv Fault]{
		\includegraphics[width=0.50\linewidth]{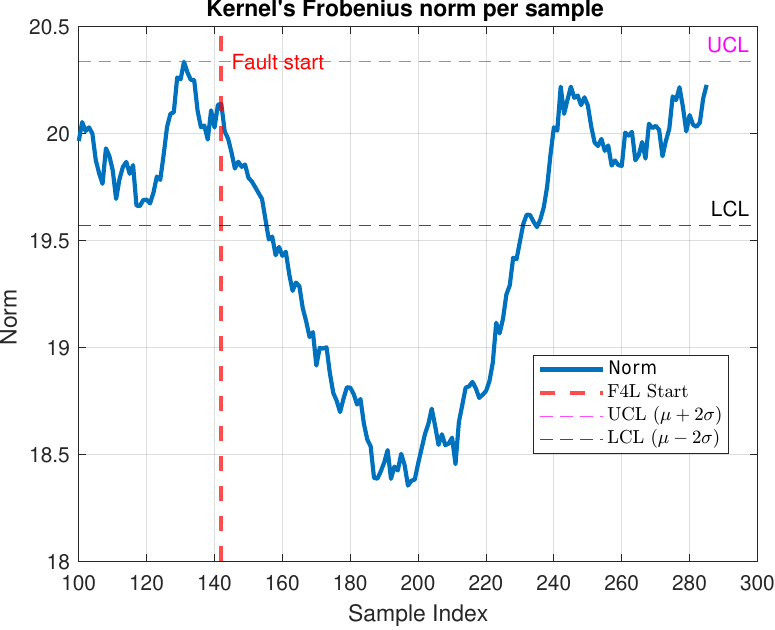}
    }
\caption{The Frobenius KNM of FFT features for individual variables under the F4L fault scenario. The blue solid line shows the norm over sliding windows (samples 100–300), with fault onset at sample 141 (vertical red dashed line). Control limits (upper: magenta, lower: black) are derived from healthy operation. Each variable exhibits a distinct fault response: (a) Ipv drops below the LCL and stabilizes to a new position; (b) Vf rises above the UCL with a non‑monotonic peak; (c) Vdc shows sustained exceedance; (d) Vpv drops sharply then returns toward the limits.}
\label{fig:F4L_FFT3}
\end{figure}

Considering next the PV array mismatch or partial shading fault (F$4$L) in Fig.~\ref{fig:F4L_FFT3}, using a similar proposed framework. From the results, we can see successful fault detection in all four variables. Once the fault is introduced, the norm exceeds the control limits in either an upward or downward direction, with zero false alarms and minimal delays, and the fault is signaled for immediate action. Among the four variables, Ipv, Vf, and Vdc seem to be more sensitive to fault detection. The approach was also evaluated for the two faults F$2$L and F$4$L when signal variables are decomposed into six levels, and the results are shown in Fig.~\ref{fig:F2L_FFT6} and Fig.~\ref{fig:F4L_FFT6} in Appendix~\ref{app:E1}. From the plots, we observe small improvements in fault detection across all fault scenarios and variables, especially for the Vf variable. For other variables, the performance remained the same as before. This demonstrates the robustness of our proposed approach at different levels of feature decomposition. These results suggest that further research may be needed to develop an efficient optimization method to determine the optimal signal decomposition level.

\begin{figure}[H]
	\centering
	\subfloat[F2L combined]{
		\includegraphics[width=0.49\linewidth]{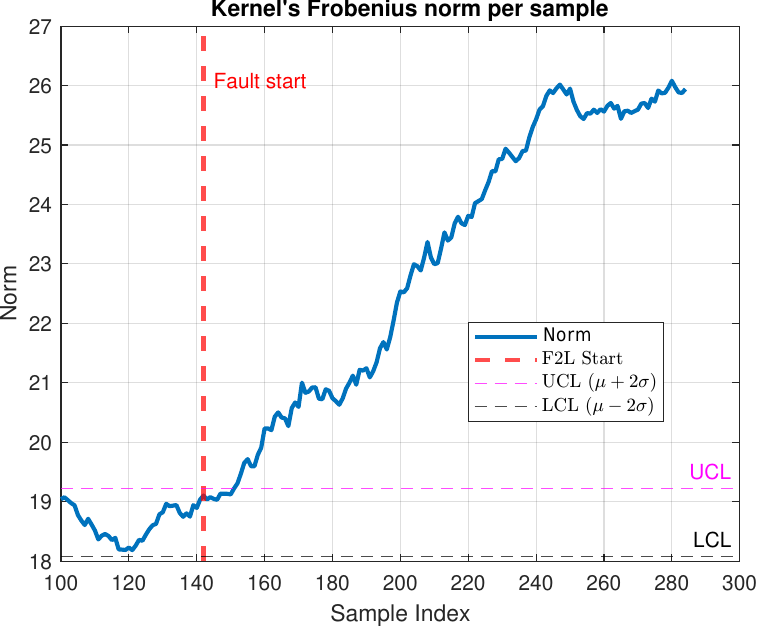}
	}
	\subfloat[F4L combined]{
		\includegraphics[width=0.50\linewidth]{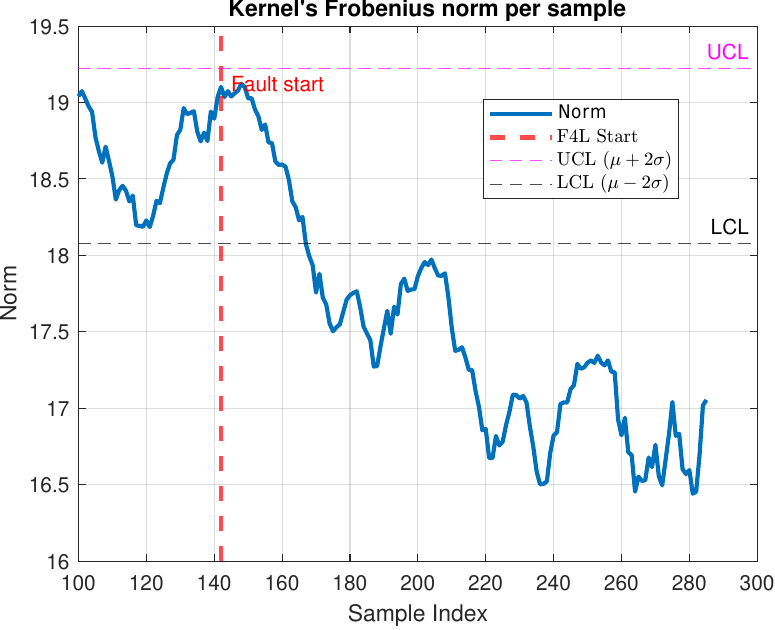}
	}\
    \subfloat[F2L variable contribution]{
    \label{Figs/Varcontr_F2}
		\includegraphics[width=0.49\linewidth]{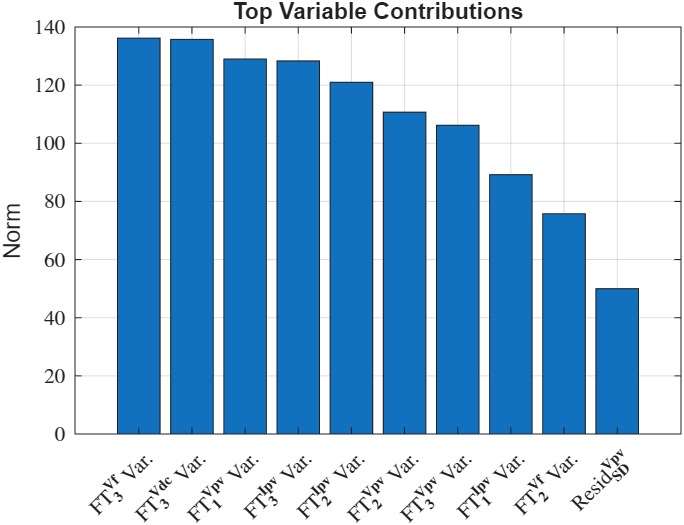}
	}
     \subfloat[F4L variable contribution]{
     \label{Figs/Varcontr2_F4}
    
		\includegraphics[width=0.49\linewidth]{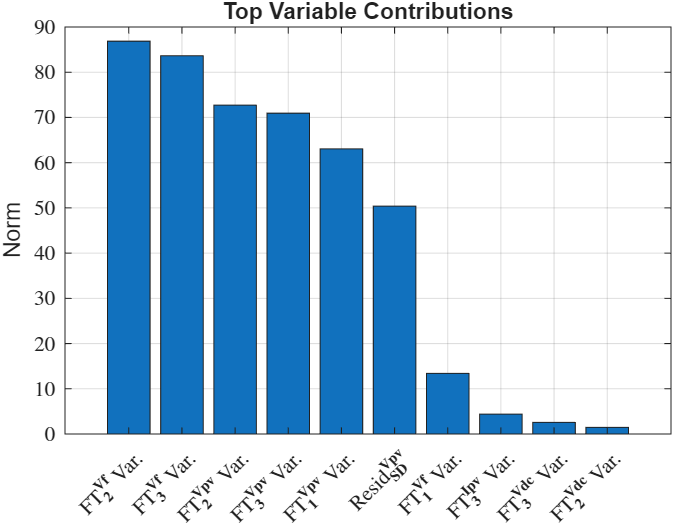}
    }
\caption{Kernel's Frobenius norm monitoring over time for a combined 3-decomposition level FFT variable features under the two fault scenarios and their contribution plots. (a)Top-left: results for the combined ($v \in \{\text{Ipv},\text{Vpv},\text{Vdc},\text{Vf}\}$) FFT features under the F2L fault scenario, where the norm increases sharply above the UCL, a few samples after the fault onset at 141 samples, providing a clear fault detection. (b)Top-right: results for the combined FFT features under the F4L fault scenario, where the norm decreases consistently below the LCL, some samples later. (c)Bottom-left: Contributions of the F2L scenario to $m_{t_k}$. (d)Bottom-right: Contributions of the F4L scenario to $m_{t_k}$. In the contribution plots, labels of the form "$\text{FT}_{i}^{v}\text{Var}.$" correspond to the variance feature $\text{V}_{i,t}^{v}$ discussed in Section~\ref{sec:pre-treatment}, whereas "$\text{Resid}_{SD}^{v}$" denotes the residual feature $\sigma(r_t^{v})$.}
\label{fig:F24L_FFT3}
\end{figure}

Further improvements in F$2$L fault detection can be seen in Fig.~\ref{fig:F24L_FFT3}, after a combination of all features extracted under a three-level FFT decomposition of the four variables used in this study. We can observe a more stable norm once the fault is detected after crossing the CL, compared to the previous analysis. Contribution plots for fault diagnosis in the combined-feature case are shown in Figs.~\ref{Figs/Varcontr_F2} and \ref{Figs/Varcontr2_F4}, respectively. These plots reveal that Vdc and Vpv are the most informative variables for both fault scenarios, whereas the remaining variables contribute to a lesser extent. This finding indicates that sensor and partial shading faults predominantly disturb the DC-link voltage regulation and PV-side voltage, explaining the strong fault detection performance achieved by the proposed monitoring framework reported in Table~\ref{tab:Mult_detection_metrics} for both faults.

\subsection{Analysis of fault detection performance results under univariate and multivariate cases}

Tables~\ref{tab:Univ_detection_metrics} and \ref{tab:Mult_detection_metrics} show the performance of our proposed approach with other models(iForest, LOF, and OCSVM) across two fault scenarios in IPPT mode under univariate and multivariate cases. For the feedback sensor fault (F2L) in the univariate case, our proposed approach excelled in FAR, precision, and AUC metrics, achieving values of (0.0465, 0.9859, 0.9486), respectively, at different kernel functions. iForest showed the highest accuracy and precision at 0.9620 and 0.9759 values. LOF followed with an accuracy of 0.9565 and an F1-score of 0.9726. For the partial shading fault (F4L), Ours (RBF) excelled at all metrics (FAR = 0.0465, accuracy = 0.9516, precision = 0.9855, F1-score = 0.9680, and AUC = 0.9523) except for the recall value, followed by iForest (FAR = 0.1667, accuracy = 0.9459, precision = 0.9524, Recall = 0.9790, and F1-score = 0.9655), then LOF (FAR = 0.1905, accuracy = 0.9189, precision = 0.9444, recall = 0.9510, and F1-score = 0.9477), and lastly by OCSVM. The results show the robustness of the proposed framework at different kernel functions.

Conversely, in the multivariate case for fault F2L, our proposed approach achieved near-perfect detection performance with an F1-score of 97\% and zero FAR, indicating strong sensitivity and precision in capturing discrete, localized signal deviations. This is followed by an LOF F1-score of 95\% and FAR = 0.1. iForest and OCSVM then followed closely. OCSVM is the last, reflecting the difficulty in identifying anomalies involving partial signal loss. In F4L, which introduces moderate non-stationarity and inter-panel mismatch, was best detected by LOF (F1-score = 96\%) and iForest (F1-score = 95\%). Followed by OCSVM (F1-score = 96\%) and ours (F1-score = 91\%). Followed by OCSVM (F1-score = 93\%) and our proposed approach (F1-score = 91\% ). This underscores the limitations of OCSVM in nonlinear, diffuse fault conditions.

\begin{table}[H]
\centering
\caption{Results of detection efficiency of the proposed method (Univariate-Vdc) and different unsupervised models with combined FFT features for F2L and F4L faults}
\label{tab:Univ_detection_metrics}
\begin{tabular}{@{}ccccccccc@{}}
\toprule
\textbf{Faults} & \textbf{Model} & \textbf{FAR} & \textbf{Accuracy} & \textbf{Precision} & \textbf{Recall} & \textbf{F1-Score} & \textbf{AUC} \\
\midrule
F2L & iForest & 0.1667  & \textbf{0.9620} & 0.9530 & 1 & \textbf{0.9759} & 0.9167  \\
F2L & LOF & 0.1905 & 0.9565 & 0.9467 & 1 & 0.9726 & 0.9048 \\
F2L & OCSVM & 0.2381 & 0.9457 & 0.9342 & \textbf{1} & 0.9660 & 0.8810 \\
F2L & Ours(RBF) & \textbf{0.0465} & 0.9459 & \textbf{0.9859} & 0.9437 & 0.9640 & \textbf{0.9486} \\
F2L & Ours(Mat{\'e}rn) & \textbf{0.0465} & 0.9351 & 0.9851 & 0.9296 & 0.9565 & 0.9416 \\
F2L & Ours(Cauchy) & \textbf{0.0465} & 0.9459 & 0.9853 & 0.9437 & 0.9640 & \textbf{0.9486} \\
\addlinespace
F4L & iForest & 0.1667 & 0.9459 & 0.9524 & 0.9790 & 0.9655 & 0.9062 \\
F4L & LOF & 0.1905 & 0.9189 & 0.9444 & 0.9510 & 0.9477 & 0.8803 \\
F4L & OCSVM & 0.2381 & 0.9405 & 0.9342 & \textbf{0.9930} & 0.9627 & 0.8775 \\
F4L & Ours(RBF) & \textbf{0.0465} & \textbf{0.9516} & \textbf{0.9855} & 0.9510 & \textbf{0.9680} & \textbf{0.9523} \\
F4L & Ours(Mat{\'e}rn) & \textbf{0.0465} & 0.9247 & 0.9850 & 0.9161 & 0.9493 & 0.9482 \\
F4L & Ours(Cauchy) & \textbf{0.0465} & 0.9247 & 0.9850 & 0.9161 & 0.9493 & \ 0.9482
\\

\bottomrule
\end{tabular}
\end{table}
Table \ref{tab:Mult_detection_metrics} summarizes the detection performance of the proposed kernel-based monitoring framework and three benchmark unsupervised models implemented with scikit-learn: iForest, LOF, and OCSVM. The models were evaluated using combined multivariate FFT features derived from photovoltaic signals under two fault scenarios, namely F2L and F4L.

For the F2L fault case, the proposed kernel monitoring approach outperforms the baseline models. While iForest and LOF achieve high recall (1.0), their false alarm rates remain relatively high at 0.4762 and 0.2857, respectively. Similarly, OCSVM produces the highest FAR (0.7619), indicating poor discrimination between normal and faulty operating conditions. In contrast, the proposed method with the Radial Basis Function Kernel achieves zero false alarms and an accuracy of 0.9514, while maintaining high precision and recall values. Even stronger performance is observed when alternative kernels are used within the proposed framework. The Mat{\'e}rn Kernel ( $\nu$ = $\frac{3}{2}$) improves accuracy to 0.9730 and achieves an F1-score of 0.9822, whereas the Cauchy Kernel provides the highest detection performance, reaching an accuracy of 0.9910 and an F1-score of 0.9930 with only minimal false alarms (FAR = 0.0233).

A similar pattern is observed for the F4L fault scenario. Among the baseline models, LOF achieves the best performance, with an accuracy of 0.9351 and an F1-score of 0.9597, while iForest and OCSVM exhibit higher false alarm rates. In comparison, the proposed method consistently provides competitive results. In particular, the Cauchy kernel again achieves the best performance with an accuracy of 0.9830 and an F1-score of 0.9857, indicating robust detection capability across different fault conditions.

These results highlight two key observations. First, the proposed kernel monitoring framework significantly reduces false alarms compared with conventional unsupervised detectors. Second, the detection performance remains consistently high across different kernel functions, demonstrating the robustness of the proposed methodology with respect to kernel selection.
\begin{table}[H]
\centering
\caption{Results of detection efficiency of the proposed method (Multivariate) and different unsupervised models with combined FFT features for F2L and F4L faults}
\label{tab:Mult_detection_metrics}
\begin{tabular}{@{}ccccccccc@{}}
\toprule
\textbf{Faults} & \textbf{Model} & \textbf{FAR} & \textbf{Accuracy} & \textbf{Precision} & \textbf{Recall} & \textbf{F1-Score} & \textbf{AUC} \\
\midrule
F2L & iForest & 0.4762 & 0.8913 & 0.8765 & 1.0 & 0.9342 & 0.7619 \\
F2L & LOF & 0.2857 & 0.9348 & 0.9221 & 1.0 & 0.9595 & 0.8572 \\
F2L & OCSVM & 0.7619 & 0.8261 & 0.8161 & 1.0 & 0.8987 & 0.6191 \\
F2L & Ours(RBF) & \textbf{0} & 0.9514 & \textbf{1} & 0.9366 & 0.9673 & 0.9683 \\
F2L & Ours(Mat{\'e}rn) & 0.0233 & 0.9730 & \textbf{0.9928} & 0.9718 & 0.9822 & 0.9743 \\
F2L & Ours(Cauchy) & 0.0233 & \textbf{0.9910} & 0.9860 & \textbf{1} & \textbf{0.9930} & \textbf{0.9849} \\
\addlinespace
F4L & iForest & 0.4762 & 0.8919 & 0.8773 & 1.0 & 0.9346 & 0.7619 \\
F4L & LOF & 0.2857 & 0.9351 & 0.9226 & 1.0 & 0.9597 & 0.8572 \\
F4L & OCSVM & 0.7619 & 0.8270 & 0.8171 & 1.0 & 0.8994 & 0.6191 \\
F4L & Ours(RBF) & \textbf{0} & 0.871 & \textbf{1} & 0.8322 & 0.9084 & 0.9161 \\
F4L & Ours(Mat{\'e}rn) & 0.0233 & 0.8925 & 0.9920 & 0.8671 & 0.9254 & 0.9219 \\
F4L & Ours(Cauchy) & \textbf{0} & \textbf{0.9830} & \textbf{1} & 0.9718 & \textbf{0.9857} & \textbf{0.9849} \\

\bottomrule
\end{tabular}
\end{table}

\subsection{Further discussion on advantages and limitations}
\label{sec:limitations}
The proposed unsupervised kernel-based change detection framework addresses a crucial gap in industrial process control by eliminating the need for costly fault-state simulations. By using a moving-window approach to estimate optimal kernel parameters, the framework provides a dynamic mechanism to track variations in nonlinear processes that classical anomaly detection schemes often miss.
The Kernel estimation method presented in this paper has the following advantages when compared to the existing literature:
\begin{itemize}
    \item I. \textit{Sensitivity and parameter optimization.}The core strength of the framework lies in its ability to optimize the kernel parameter ($\sigma_{\text{opt}}$) over a grid $\Gamma$ using solely normal functioning process data. This allows the model to learn the intrinsic geometry of the healthy state in a high-dimensional feature space without parametric restrictions. The use of the Frobenius norm of the centered kernel matrix ($m{t_k}= \|\mathbf{K}_c(\sigma_{\text{opt}})\|_F$) as a monitoring statistic effectively captures changes in the data's structure within the feature space. As highlighted in other related studies, such kernelized approaches are superior for handling non-linear features where input variables do not follow a Gaussian distribution.
    \item II. \textit{Thresholding and robustness.} The framework currently utilizes a baseline-derived threshold ($[\mu - \kappa\hat{\sigma}, \mu + \kappa\hat{\sigma}]$) to identify anomalies. While effective for initial detection, the sources suggest that real-world measurements, particularly in PV systems, often exhibit non-Gaussian, multimodal distributions. Therefore, a significant possible improvement would be to replace the parametric threshold with a nonparametric threshold derived from kernel density estimation (KDE). Utilizing KDE to set control limits based on the actual ($1-\alpha$) quantile of the decision statistic has been shown to reduce false alarm rates and increase flexibility in unpredictable environments. 
    \item III. \textit{Adaptive model updating.} Currently, the framework selects an optimal $\sigma_{\text{opt}}$ based on an initial baseline window. However, complex processes often exhibit evolving normal behavior due to factors such as aging or seasonal shifts ~\cite{sohani2022using}. To enhance long-term reliability, the algorithm could be improved by integrating an adaptive update mechanism. Similar to the discrimination index (AD) found in related studies ~\cite{bakdi2021real}, the system could trigger a re-optimization of kernel parameters when a slow, persistent shift in the mean is detected that does not correspond to a sudden fault. This would prevent "model obsolescence" and ensure the detector remains sensitive to incipient anomalies while remaining robust to natural process evolution.
    \item IV. \textit{Fault detection performance comparison}. While the proposed framework performed well with reduced variables under IPPT mode, it will be more informative to also compare its performance under MPPT mode and with other GPVS-Faults that were not validated in this study due to space limitations. This dual-mode assessment will showcase how different operating regimes influence detection performance under different fault scenarios and establish a consistent benchmark for GCPVS monitoring.
\end{itemize}

\newpage
\section{Conclusions and future work}
\label{sec:conclusion}
This study proposed kernel-based norm monitoring (KNM), a fully unsupervised fault detection framework for continuous multivariate processes operating under conditions where labeled fault data are unavailable. The framework combines DFT feature extraction, unsupervised kernel parameter optimization via variance maximization of the centered kernel matrix, and a Frobenius norm monitoring statistic tracked over a sliding window, with a per-variable contribution analysis layer for interpretable fault identification. Applied to GCPVS operating in intermediate power point tracking (IPPT) mode --- a setting characterized by compact feature distributions and reduced short-term dynamics that preclude direct transfer of models developed for maximum power point tracking conditions --- the proposed framework was evaluated against three established unsupervised benchmark methods: OCSVM, iForest, and LOF.

The results demonstrate that KNM achieves up to 99.1\% and 98.3\% detection accuracy for sensor faults and partial shading faults respectively using the Cauchy kernel, compared to 93.5\% for the best-performing benchmark method. Critically, these results are obtained without any labeled fault data at any stage of training or parameter selection, relying exclusively on normal operation data from the initial baseline window. The structured synthetic evaluation further demonstrated that the Frobenius norm statistic is sensitive to all kernel types at different fault magnitudes. The per-variable contribution plots consistently identified the variables most responsible for detected anomalies, providing physically interpretable diagnostic outputs that address a known limitation of black-box unsupervised detectors.

Future work will focus on addressing the limitations presented in Section \ref{sec:limitations} through online kernel re-calibration using adaptive sliding reference windows to accommodate process drift; evaluation across all fault categories and operating modes in the GPVS-Faults dataset; and validation on field-collected data from operational GCPVS installations, where environmental variability exceeds that of the hardware-in-the-loop emulator used here. More broadly, the proposed framework is not specific to photovoltaic systems; the combination of sliding window KNM with unsupervised parameter optimization applies to any continuous multivariate process where normal operation data is available and labeled fault data is not, including rotating machinery, chemical processes, and power electronics monitoring.

\section{Declarations}

\subsection{\textbf{Competing interests}}
The authors declare that they have no competing interests.

\subsection{\textbf{Funding}}
We acknowledge funding for the Flagship of Advanced Mathematics for Sensing, Imaging, and Modeling 2024--2031 (decision number 359183). VJ was supported through the Higher Education for Economic Transformation (HEET) program, funded by the World Bank through the Government of Tanzania.

\subsection{\textbf{Authors' contributions}}
Conceptualization: VJ, and ZD, methodology: VJ, and ZD, software: VJ and ZD, validation: VJ, ZD, HH, SR, and LR, formal analysis: VJ, and ZD, investigation: VJ, and ZD, resources: SR and LR, data curation: VJ, writing – original draft preparation: VJ, writing – review and editing: VJ, ZD, HH, and LR, visualization: VJ, and ZD, supervision: ZD, HH and LR, project administration: HH and LR, funding acquisition: SR and LR. 

\appendix

\section{Grid-connected PV system}\label{app:stft}
\begin{figure}[H]
    \centering
    \includegraphics[width=0.9\linewidth]{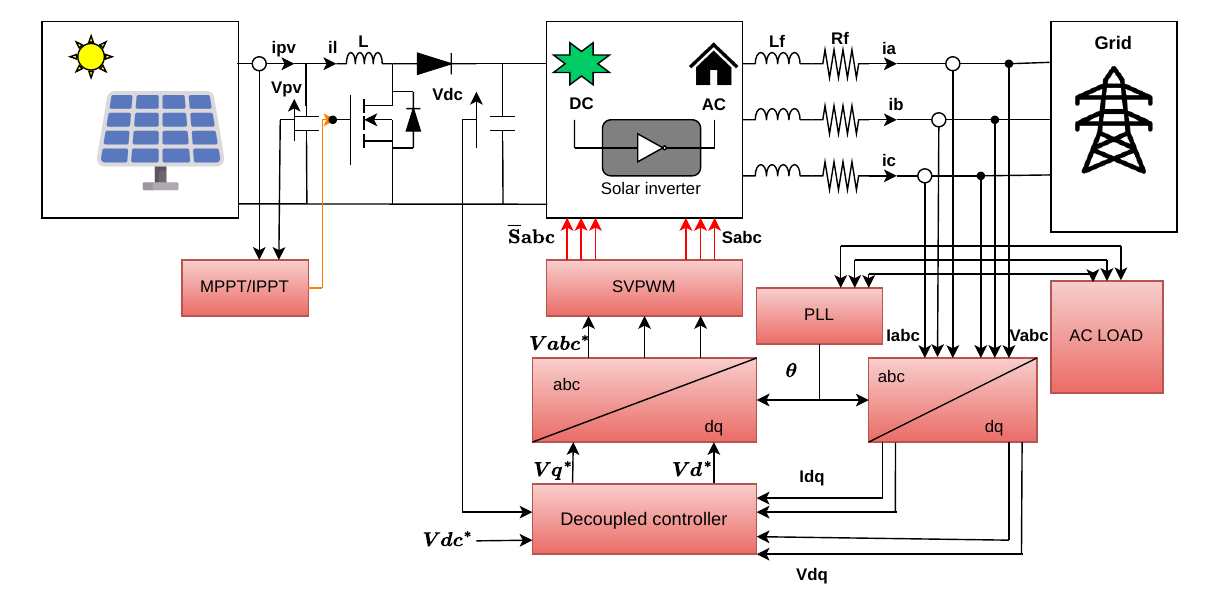}
    \caption{Block Diagram of grid-connected PV system}
    \label{fig:Blockdiag}
\end{figure}

The block diagram of the GCPVS utilized in the work is shown in 
Fig.~\ref{fig:Blockdiag}.

\section{IPPT mode exploratory data analysis}
\label{app:variables}
\begin{figure}[H]
    \centering
    \includegraphics[width=1\linewidth]{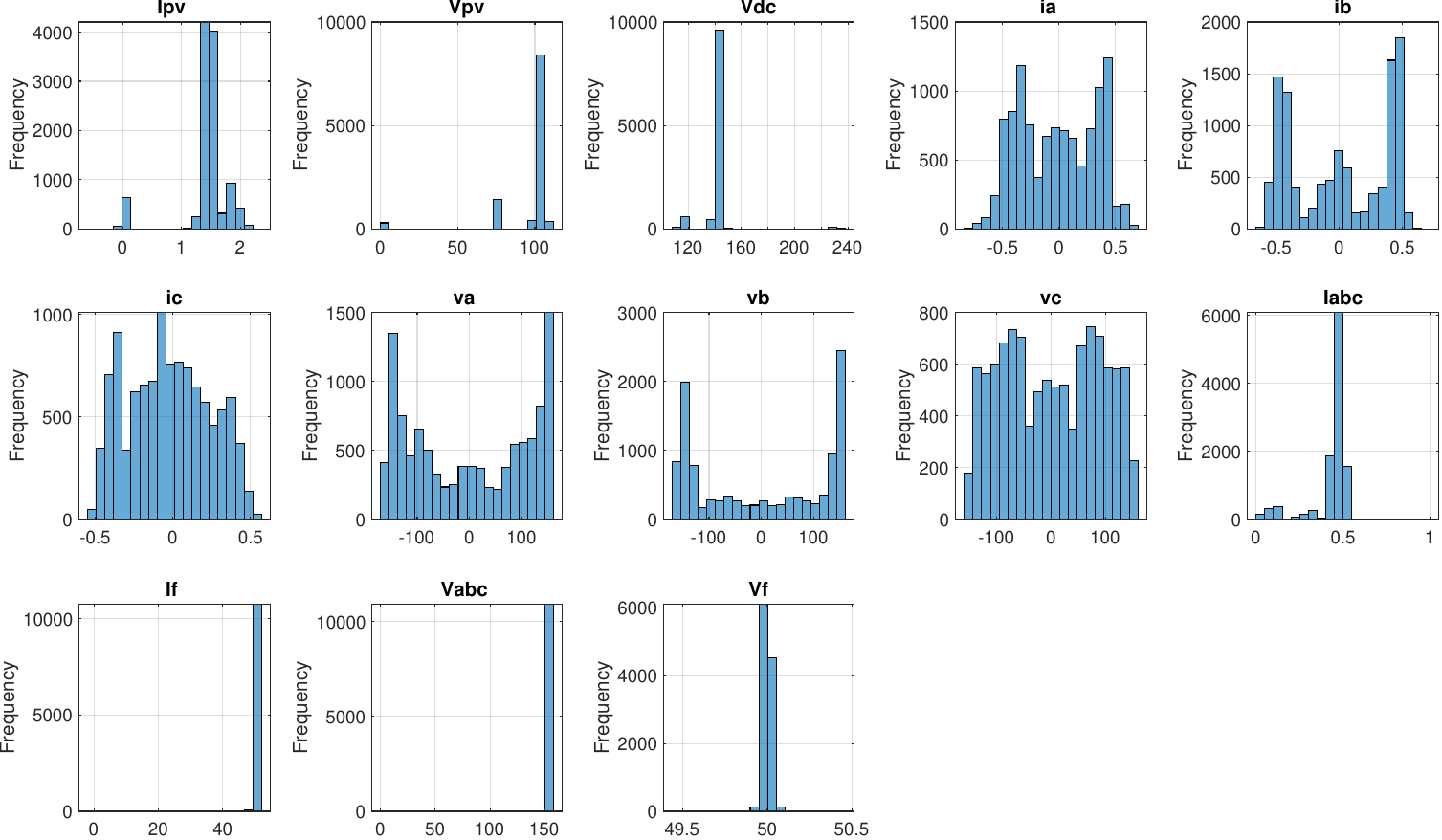}
    \caption{Distribution of healthy GCPVS variables used for multivariate statistical process monitoring. The histograms show the baseline behavior of the measured signals and illustrate the non-Gaussian and variable-specific characteristics that motivate the use of kernel-based monitoring approaches.}
    \label{fig:hist}
\end{figure}

\begin{figure}[H]
    \centering
    \subfloat[]{\label{fig:heat1}%
        \includegraphics[width=0.49\linewidth]{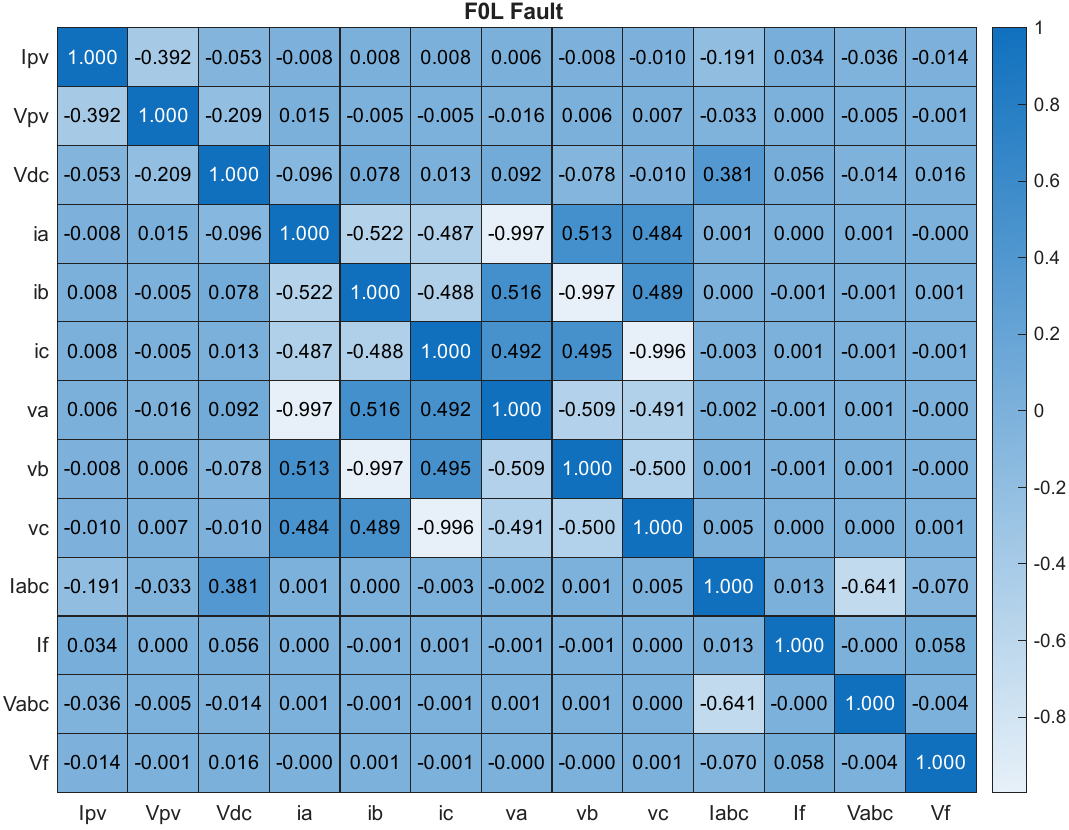}%
    }
    \hfill
    \subfloat[]{\label{fig:heat2}%
        \includegraphics[width=0.49\linewidth]{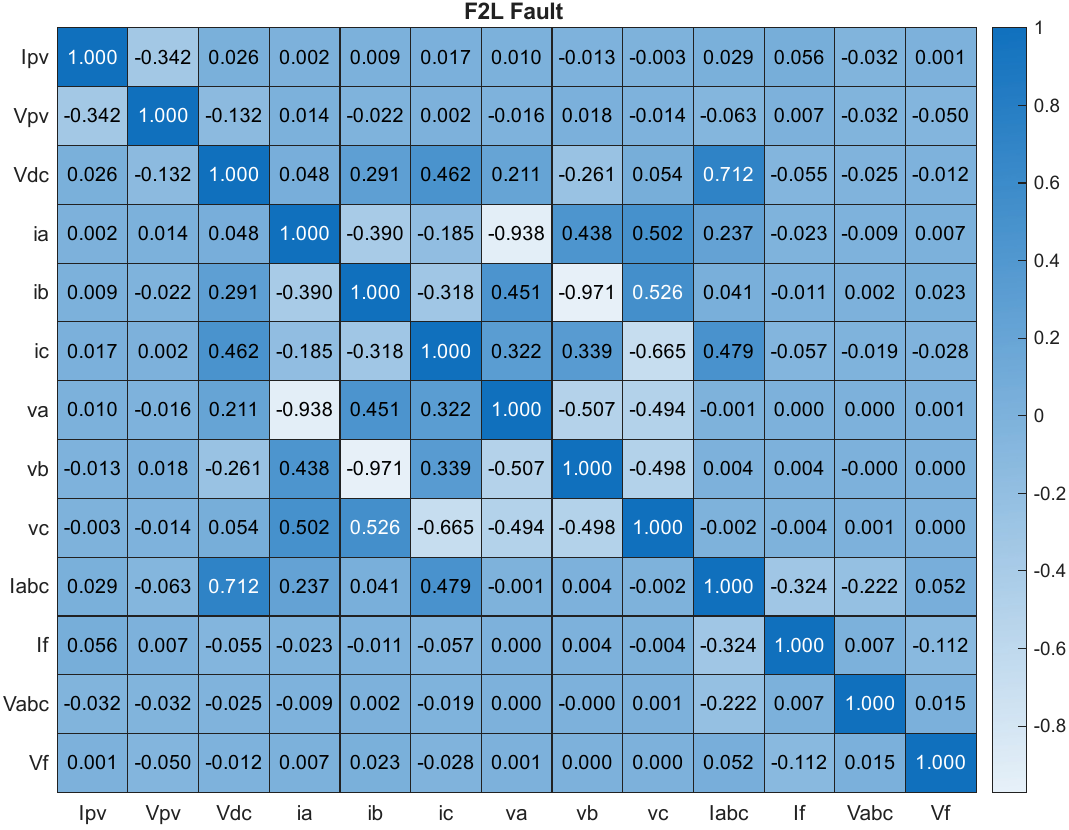}%
    }
    \caption{Pearson correlation matrices of the GPVS-Faults dataset variables under healthy (F0L) and faulty (F2L) operating conditions during IPPT mode.}
    \label{fig:heat}
\end{figure}

Fig.~\ref{fig:hist} shows the histogram distributions for the 13 data variables considered in this study in IPPT mode. The distribution plots illustrate the non-Gaussian and multimodal nature of these datasets. This non-Gaussian characteristic may pose a challenge for approaches that assume Gaussian process variables. The heatmap in Fig.~\ref{fig:heat} visually shows the pairwise correlations of the variables under intermediate-power conditions, both without faults (F0L) and with fault 2 (F2L). The correlation coefficients ranged from $-1$ to $1$. High positive or negative values signify strong correlations, while values close to 0 imply weak or no correlation.

Understanding these relations is significant because they describe correlations that can provide critical insight into fault conditions. Consequently, a strong correlation between Vpv and Ipv is expected during fault-free conditions. Adherence to this predicted relationship serves as an indicator of fault absence, while any measurable deviation constitutes evidence of the presence of a fault. This approach significantly enhances the interpretability of the relations of these variables, which is vital for developing preventive fault detection algorithms by establishing patterns indicative of abnormalities.

\section{Effect of fault magnitude on Cauchy and Mat{\'e}rn kernel norm curves}
\label{ssec:other kernels}
\begin{figure}[H]
    \centering
    \subfloat[]{
    \label{fig:kernels_case3}%
        \includegraphics[width=0.55\linewidth]{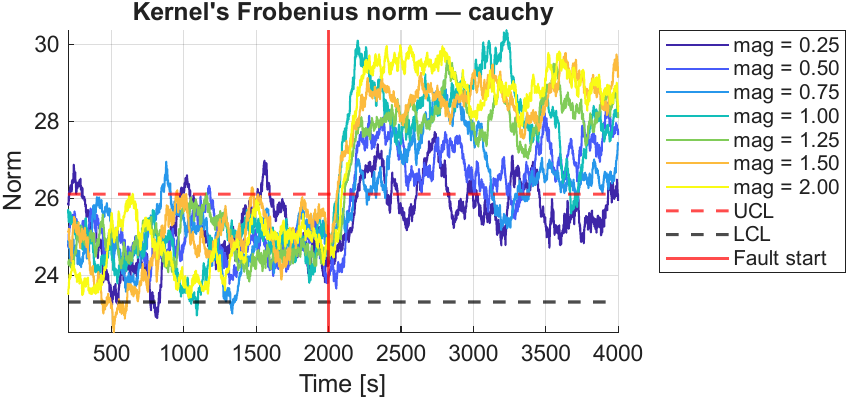}%
    }
    \hfill
    \subfloat[]{\label{fig:kernels_case2}%
        \includegraphics[width=0.55\linewidth]{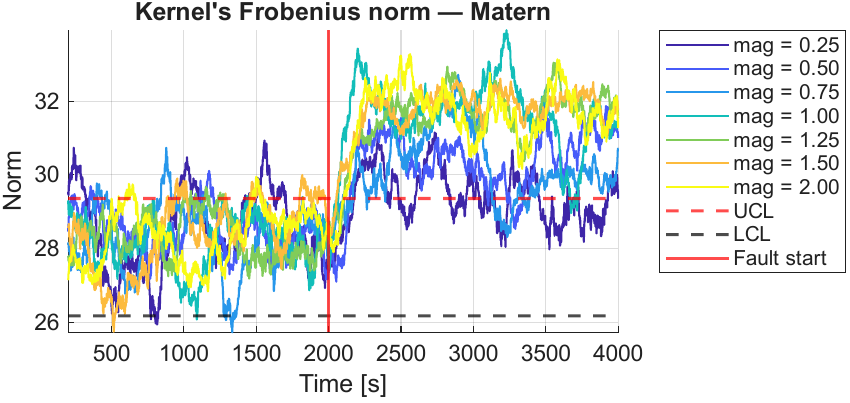}%
    }
    \caption{Cauchy kernel (a) and  Mat{\'e}rn kernel (b) Frobenius norm curves across simulated fault magnitudes. The upper plot exhibits heavier-tailed response behavior, with more gradual departures at lower magnitudes and sharper excursions at higher magnitudes. While the lower plot produces a smoother norm evolution, with moderate sensitivity to fault size.}
\end{figure}

Fig.~\ref{fig:kernels_case3} shows the same experiment as previously discussed for Fig.~\ref{fig:simg_rbf}, here with different kernels (Cauchy and Mat{\'e}rn). Among the investigated kernels, differences in sensitivity were observed. The RBF kernel generally produced sharper separations between healthy and faulty conditions, reflecting its strong ability to capture localized non-linear changes. The Cauchy kernel exhibited a more gradual response consistent with its heavier-tailed similarity function, while a Mat{\'e}rn kernel provided an intermediate behavior that balances smoothness and robustness. These differences suggest that kernel choice influences fault sensitivity and should be considered as an important design parameter in kernel-based monitoring systems.

\section{Experimental setup and parameter configuration for the benchmark models}
This subsection describes the experimental design and parameter settings used to evaluate the effectiveness of our proposed framework and three other semi-supervised anomaly detection methods, iForest, LOF, and OCSVM, for monitoring GCPVS operating under IPPT mode. Where IPPT is commonly employed when power generation is constrained relative to demand, resulting in regulated
operation at intermediate power levels.
Faults arising in this mode can reduce efficiency, lead to power losses, or cause unsafe operating conditions if left undetected. The models were trained primarily on fault-free data window segments to learn normal operating behavior, which is relevant to realistic deployment scenarios in which labeled fault data are scarce. All methods were configured through systematic hyperparameter optimization via grid search, with model selection based on validation performance using the selected evaluation metrics. Table~\ref{tab:hyperparameters} summarizes the final configurations used in the experiments. For Isolation Forest, a random\_state value of 42 and contamination of 0.1 yielded the best performance, and the other parameters followed their default configurations. For LOF, the number of
neighbors was optimized over the range [10, 20], with 15 selected as
optimal, while the remaining parameters were kept at their default values. For OCSVM, an RBF kernel was adopted; the $\nu$ parameter was tuned in the range [0.01, 0.1], with 0.1 selected as optimal, and gamma was fixed to the default setting ( $\sigma$ = scale), while the other parameters were kept at their default values. The models were evaluated on a dataset comprising both normal and fault-induced scenarios, including measurements of voltage, current, frequency, and power. The evaluation focuses on each method's ability to detect faults in the presence of noise and multimodal behavior. The analysis focuses on the IPPT regime to assess the robustness, sensitivity, and practical suitability of the proposed technique relative to other unsupervised models.

\begin{table}[H]
\centering
\caption{Hyperparameters used for each baseline model in our experiments.}\label{tab:hyperparameters}
\begin{tabular}{l @{\hspace{15pt}} l @{\hspace{20pt}} l}
\toprule
Model    & Parameter         & Value \\
\midrule
\multirow{2}{*}{iForest} & contamination   & 0.1 \\
                         & random\_state      & 42 \\
\midrule
\multirow{2}{*}{LOF} & contamination    & 0.1 \\
                     & Novelty          & True \\ 
                     &n\_neighbors
                             &
                     15\\        
\midrule
\multirow{3}{*}{OCSVM} & Kernel          & RBF \\
                      & $\nu$              & 0.1 \\
                      & $\gamma$           &
                      scale\\
\bottomrule
\end{tabular}
\end{table}


\section{More results on the proposed approach}
\subsection{Higher decomposition level on FFT features}
\label{app:E1}
\begin{figure}[H]
	\centering
	\subfloat[Ipv FFT features]{
        \label{Figs:F2L_Ipv6}
		\includegraphics[width=0.49\linewidth]{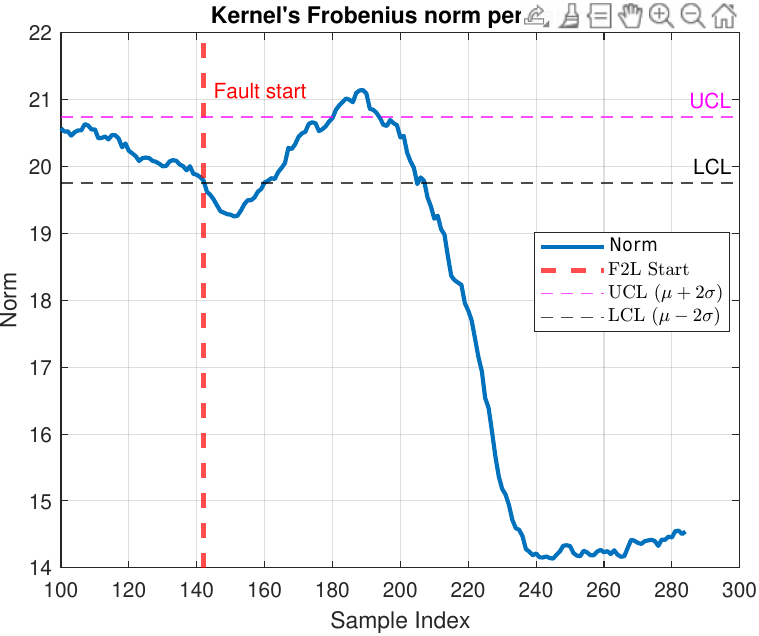}
	}
	\subfloat[Vf FFT features]{
        \label{Figs:F2L_Vf6}
		\includegraphics[width=0.49\linewidth]{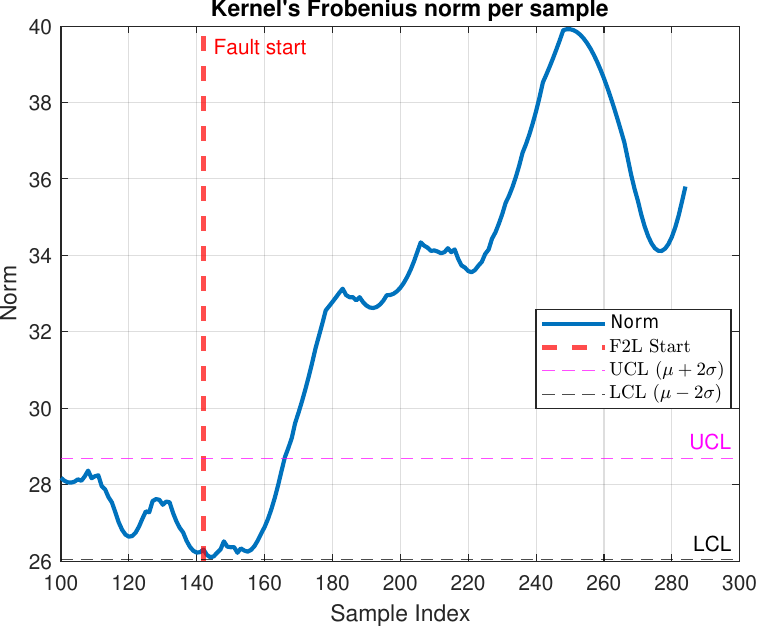}
	}\
    \subfloat[Vdc FFT features]{
        \label{Figs:F2L_Vdc6}
		\includegraphics[width=0.49\linewidth]{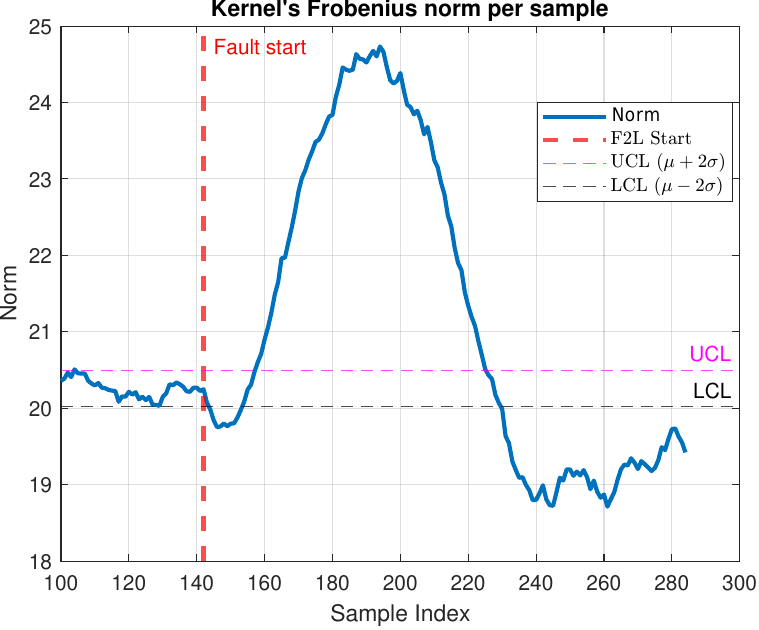}
	}
     \subfloat[Vpv FFT features]{
        \label{Figs:F2L_Vpv6}
		\includegraphics[width=0.49\linewidth]{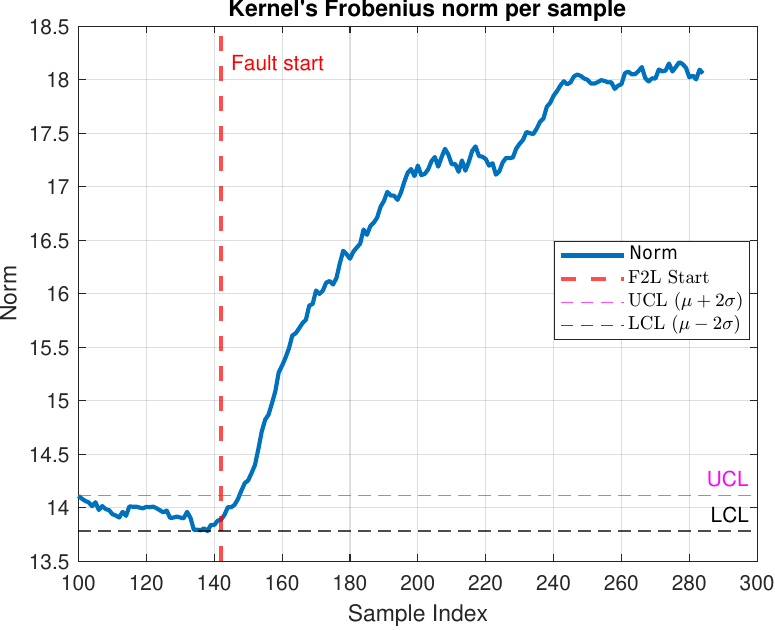}
    }
\caption{A KNM statistic based on the first six FFT spectral feature representations of four photovoltaic variables under the F2L fault condition: (a) Ipv, (b) Vf, (c) Vdc, and (d) Vpv. FFT features extracted from sliding windows are mapped into an RKHS using an RBF kernel, and the resulting centered kernel matrix is monitored through its Frobenius norm. The fault is introduced at sample 141 (vertical red dashed line), while the upper and lower control limits correspond to the $\mu \pm 2\hat{\sigma}$ limits estimated from healthy operating data.}
\label{fig:F2L_FFT6}
\end{figure}

\begin{figure}[H]
	\centering
	\subfloat[Ipv Fault]{
        \label{Figs:F4L_Ipv6}
		\includegraphics[width=0.49\linewidth]{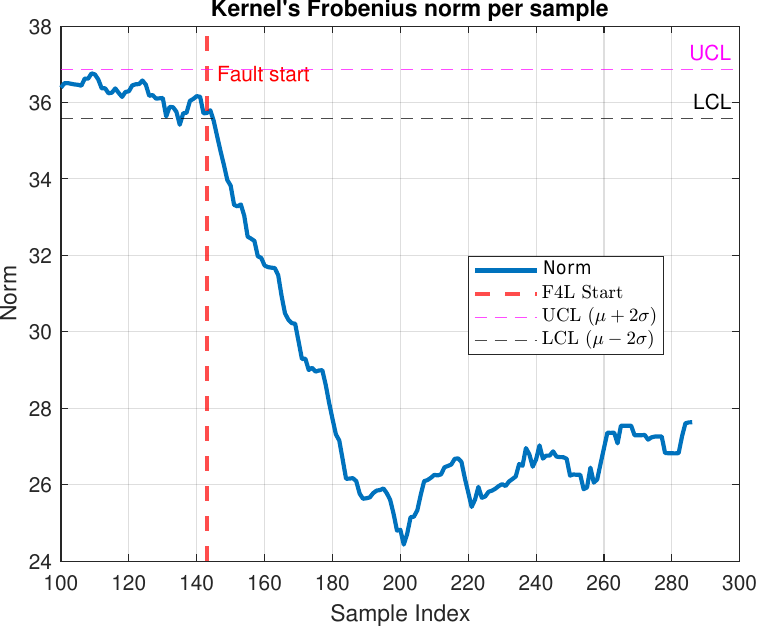}
	}
	\subfloat[Vf Fault]{
        \label{Figs:F4L_Vf6}
		\includegraphics[width=0.49\linewidth]{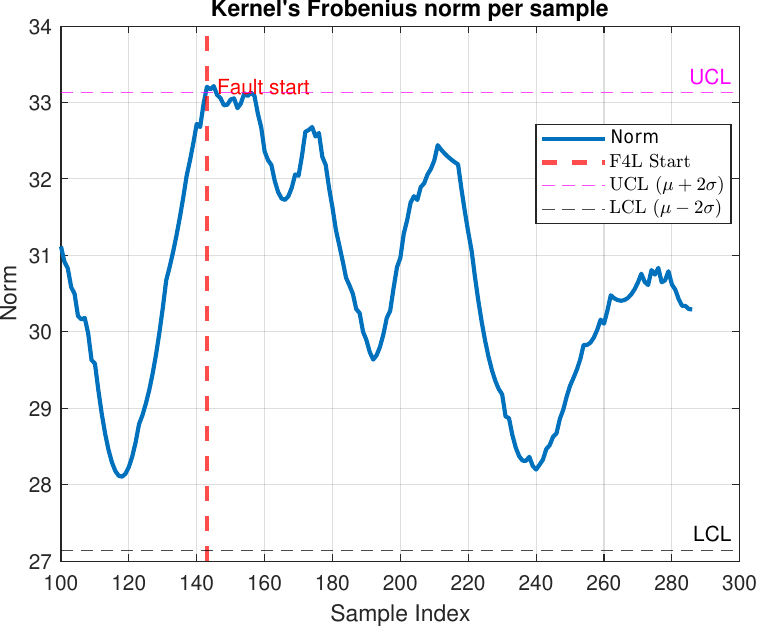}
	}\
    \subfloat[Vdc Fault]{
        \label{Figs:F4L_Vdc6}
		\includegraphics[width=0.49\linewidth]{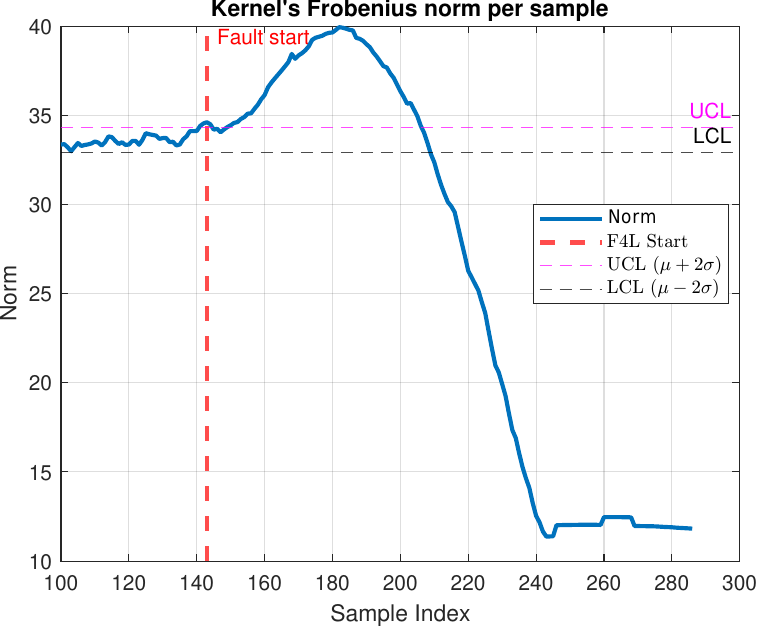}
	}
     \subfloat[Vpv Fault]{
        \label{Figs:F4L_Vpv6}
		\includegraphics[width=0.49\linewidth]{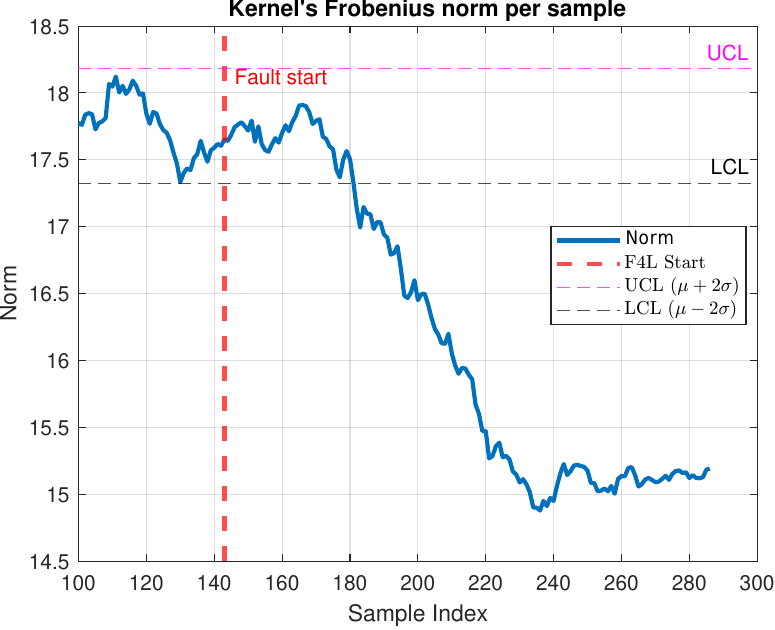}
    }
\caption{A KNM statistic based on the first six FFT spectral feature representations of four photovoltaic variables under the F4L fault condition: (a) Ipv, (b) Vf, (c) Vdc, and (d) Vpv. FFT features extracted from sliding windows are mapped into an RKHS using an RBF kernel, and the resulting centered kernel matrix is monitored through its Frobenius norm. The fault is introduced at sample 141 (vertical red dashed line), while the upper and lower control limits correspond to the $\mu \pm 2\hat{\sigma}$ limits estimated from healthy operating data.}
\label{fig:F4L_FFT6}
\end{figure}

Figs.~\ref{fig:F2L_FFT6} and ~\ref{fig:F4L_FFT6} show the KNM monitoring statistic $m_{t_k}=\|\mathbf{K}_c(\sigma_{\text{opt}})\|_F$ for the four PV variables under F2L and F4L fault conditions using six FFT spectral features. During healthy operation, the norm remains stable within the control limits across all variables. Following fault onset at sample 141 (vertical red dashed line), the statistic rises above the UCLs to a stable position in Figs.~\ref{Figs:F2L_Vf6} and \ref{Figs:F2L_Vpv6}, and drops below the LCLs in Figs.~\ref{Figs:F4L_Ipv6} and \ref{Figs:F4L_Vpv6}, with more delays in the latter plot. Moreover, the statistic proves its robustness in fault detection in other variables too, despite some delays and fluctuations after fault detection before it stabilizes to a new position. These results confirm that indeed the fault induces a structural change in the kernel feature space (more prominently for the Vpv variable for both faults) that the Frobenius norm captures reliably.

\subsection*{Appendix E.2: Multivariate unsupervised KNM fault detection: other kernel functions}
\begin{figure}[H]
	\centering
	\subfloat[]{
	    \label{Figs:F2L_cmatern}
        \includegraphics[width=0.49\linewidth]{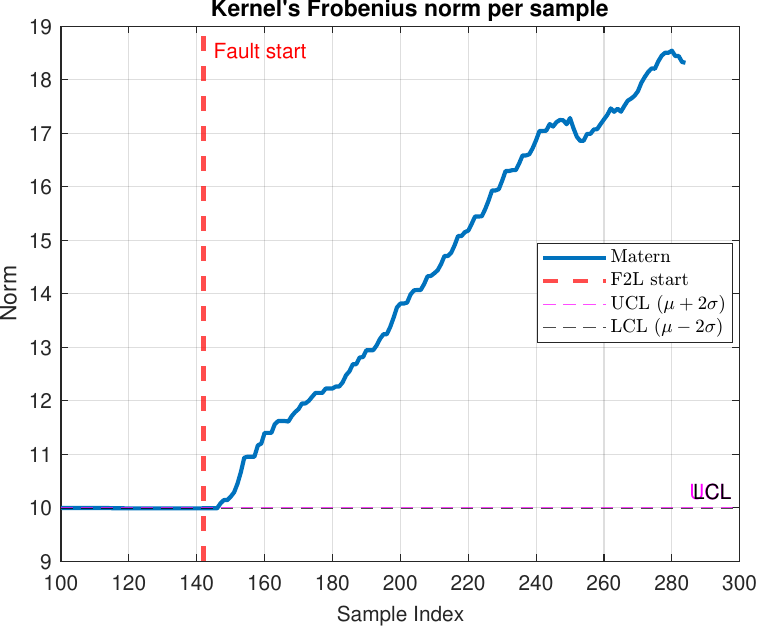}
	}
	\subfloat[]{
		\label{Figs:F2L_ccauchy}
        \includegraphics[width=0.49\linewidth]{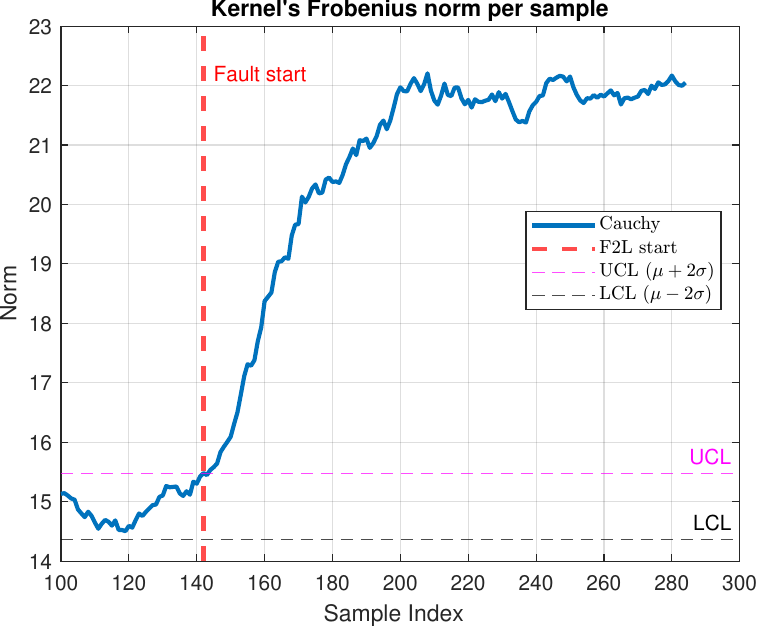}
	}\
    \subfloat[]{
		\label{Figs:F4L_cmatern}
        \includegraphics[width=0.49\linewidth]{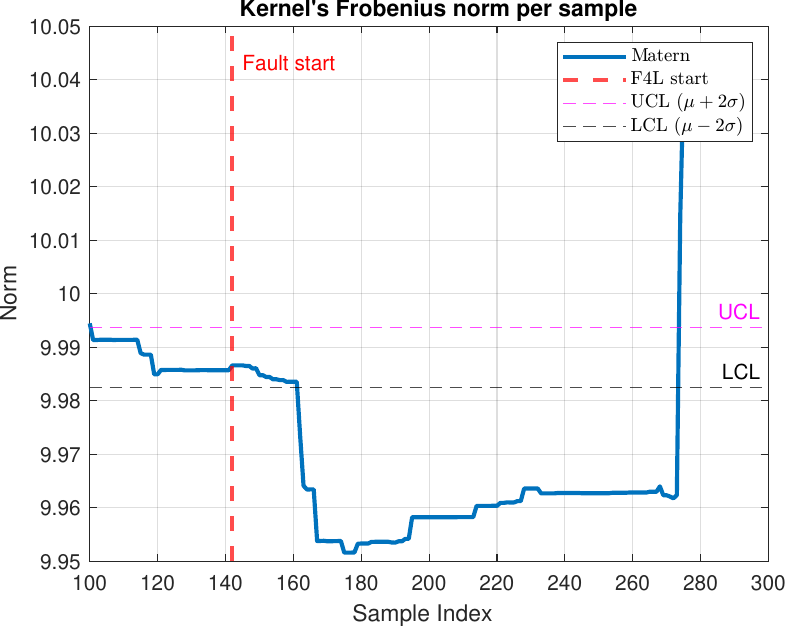}
	}
     \subfloat[]{
         \label{Figs:F4L_ccauchy}
	     \includegraphics[width=0.49\linewidth]{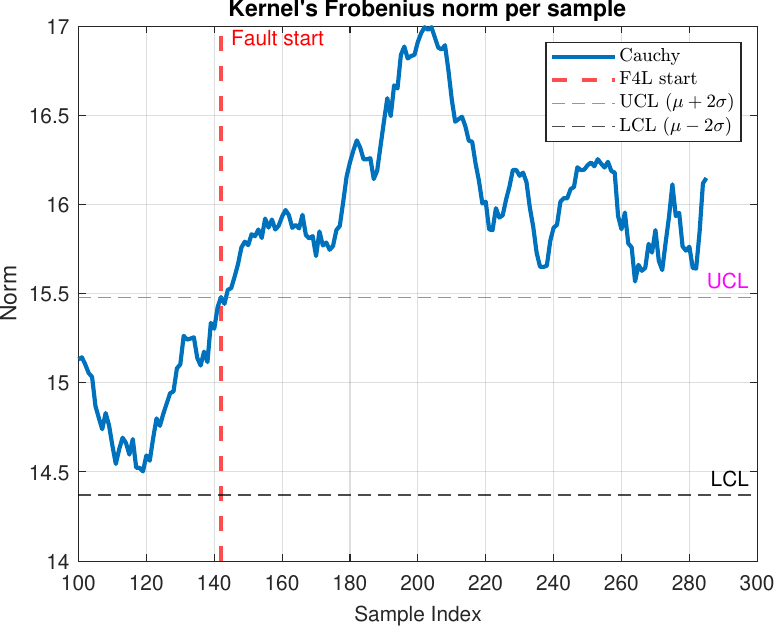}
    }
\caption{KNM statistic for the combined four photovoltaic variables (Ipv, Vpv, Vdc, Vf) using the first three FFT spectral features under F2L and F4L fault conditions:(a) F2L fault, Mat{\'e}rn $\frac{1}{2}$, (b) F2L fault, Cauchy kernel, (c) F4L fault, Mat{\'e}rn $\frac{1}{2}$, and (d) F4L fault, Cauchy kernel. The monitoring statistic $m_{t_k}=\|\mathbf{K}_c(\sigma_{\text{opt}})\|_F$ is tracked over sliding windows, with the fault injected at t=141 samples (vertical red dashed line). Control limits UCL and LCL are set at $\mu \pm 2\hat{\sigma}$ from healthy baseline data.}
\label{fig:Fcomb_Kcomp}
\end{figure}

Fig.~\ref{fig:Fcomb_Kcomp} illustrates the temporal evolution of the KNM statistic computed as the Frobenius norm of the kernel matrix under the sliding kernel parameter optimization framework. The plots show the norm trajectories for multivariate FFT features using the Matérn and Cauchy kernels, together with control limits derived from the baseline window with healthy data.

During the healthy operating window, the statistic remains stable and lies within the control limits, indicating that the similarity structure of the baseline data is consistently captured by the kernel representation. Following the fault initiation point, marked by the vertical dashed line, the monitoring statistic exhibits a clear deviation from the baseline region. This deviation reflects a change in the underlying data distribution caused by the fault, which is effectively captured in the kernel feature space. For the Mat{\'e}rn kernel, as in Fig.~\ref{Figs:F2L_cmatern} and Fig.~\ref{Figs:F4L_cmatern}, the norm trajectory increases or decreases gradually after fault onset, indicating a smooth transition from healthy to faulty operating conditions. This behavior is consistent with the controlled smoothness properties of the Mat{\'e}rn kernel. 

In contrast, the Cauchy kernel, as shown in Fig.~\ref{Figs:F2L_ccauchy} and Fig.~\ref{Figs:F4L_ccauchy} exhibits a more pronounced deviation from the baseline region, producing a stronger separation between healthy and faulty states. The heavier-tailed similarity structure of the Cauchy kernel allows it to amplify distributional changes in the multivariate spectral features, resulting in improved fault detectability. Overall, the KNM statistic demonstrates clear sensitivity to system disturbances, with both kernels providing reliable detection of fault occurrence. The consistency of these results across different kernel functions further supports the robustness of the proposed sliding kernel monitoring framework.

\newpage

\bibliographystyle{plain} 
\bibliography{Mybib.bib}

\end{document}